\documentclass[%
 aip,
 amsmath,amssymb,
 reprint,%
]{revtex4-1}
\usepackage{graphicx}
\usepackage{dcolumn}
\usepackage{bm}
\usepackage[utf8]{inputenc}
\usepackage[T1]{fontenc}
\usepackage{mathptmx}
\usepackage{etoolbox}
\usepackage[english]{babel}
\usepackage{etoolbox}
\usepackage{lipsum}
\usepackage{comment}
\usepackage{listings}
\usepackage{array}

\newcommand{\code}[1]{\lstinline!#1!}
\usepackage[table,xcdraw]{xcolor}
\usepackage[labelfont=bf]{caption}
\usepackage{subcaption}
\usepackage{hyperref}

\makeatletter
\def\@email#1#2{%
 \endgroup
 \patchcmd{\titleblock@produce}
  {\frontmatter@RRAPformat}
  {\frontmatter@RRAPformat{\produce@RRAP{*#1\href{mailto:#2}{#2}}}\frontmatter@RRAPformat}
  {}{}
}%
\makeatother
\begin{document}

\preprint{AIP/123-QED}

\title[aims-PAX: Parallel Active eXploration Enables Expedited Construction of Machine Learning Force Fields for Molecules and Materials]{aims-PAX: Parallel Active eXploration Enables Expedited Construction of Machine Learning Force Fields for Molecules and Materials}
\author{Tobias Henkes}
\affiliation{ 
Department of Physics and Materials Science, University of Luxembourg, L-1511 Luxembourg, Luxembourg
}
\author{Shubham Sharma}
\affiliation{%
Max Planck Institute for the Structure and Dynamics of Matter, 22761 Hamburg, Germany
}%
\author{Alexandre Tkatchenko}%
\affiliation{ 
Department of Physics and Materials Science, University of Luxembourg, L-1511 Luxembourg, Luxembourg
}%
\author{Mariana Rossi}
\affiliation{%
Max Planck Institute for the Structure and Dynamics of Matter, 22761 Hamburg, Germany
}%
\author{Igor Poltavskyi}%
\email{igor.poltavskyi@uni.lu}
\affiliation{ 
Department of Physics and Materials Science, University of Luxembourg, L-1511 Luxembourg, Luxembourg
}

\date{\today}

\begin{abstract}
Recent advances in machine learning force fields (MLFF) have significantly extended the reach of atomistic simulations. Continuous progress in this field requires reliable reference datasets, accurate MLFF architectures, and efficient active learning strategies to enable robust modeling of complex molecular and material systems. Here we introduce \textsc{aims-PAX}, an expedited, multi-trajectory active learning framework that streamlines the development of stable and accurate MLFFs. Designed for a wide range of researchers, \textsc{aims-PAX} offers a modular, high-performance workflow that couples diversified sampling with scalable training across CPU and GPU architectures.
Integrated with the widely used \textit{ab initio} code \textsc{FHI-aims}, the framework supports state-of-the-art ML models and dataset generation using general-purpose (or "foundational") force-fields for rapid deployment in diverse systems. We demonstrate the capabilities of \textsc{aims-PAX} in various challenging tasks: creating datasets and models for highly flexible peptides, multiple organic molecules at once, explicitly solvated molecules, and for efficiently handling computationally demanding systems such as the CsPbI\textsubscript{3} perovskite. We show that \textsc{aims-PAX} achieves a reduction of up to three orders of magnitude in the number of required reference calculations, automatically selects challenging systems within a given chemical space, facilitates simulation of solvated molecules with more than thousand atoms, while enabling a ten-fold speedup in active-learning time through optimized resource utilization. This positions \textsc{aims-PAX} as a powerful and versatile platform for next-generation atomistic simulations in both academic and industrial settings.
\end{abstract}

\maketitle


\section{\label{sec:Introduction}Main:\protect\\ }
The successes of machine learning force fields (MLFFs)\cite{mlff_review} have deeply transformed the field of molecular simulations. They are now the preferred method for simulating the dynamics of large systems, such as perovskites~\cite{baldwin2024} or solvated proteins~\cite{unke2024_GEMS}, with quantum-chemical accuracy. While general-purpose (GP) (sometimes called ``foundational'') models~\cite{Kabylda2025_so3lr,Smith2017_ANI, Kovcs2025_maceoff, Chen2019_foundational, Deng2023_foundational, Choudhary2021_foundational, Merchant2023_deepmind_foundational, gasteiger2021_foundational} trained on large datasets~\cite{Hoja2021_qm7x,Eastman2023_spice, matprj_dataset, omat24, omol25, Ganscha2025_qcml} are becoming more widespread, there remains a strong demand for high-quality data to fine-tune these models or to build new, and often cheaper, custom models for challenging applications.~\cite{novelli2024fine_finetune1, Kaur2025_finetune2, maurer2025_finetune3} 

The process of collecting representative high-quality datasets can be labor-intensive, requiring considerable manual effort and computational resources. To address these challenges, a common approach is to employ active learning (AL).~\cite{settles_active_2009, survery_DL_AL} In AL, an uncertainty measure of a model prediction is used to select data points for labeling and inclusion in the training dataset. This approach enriches the training dataset with points that represent a challenge for the current state of the model. In essence, the model autonomously determines which data to prioritize for training and which to disregard.
Therefore, this procedure reduces human intervention and decreases the computational cost of model training by requiring only a small number of expensive and slow reference \textit{ab initio} calculations to reach an acceptable accuracy. In addition, AL also improves the robustness of the MLFF by detecting and correcting possible failures during the training procedure.

AL has been successfully applied to a plethora of applications. For example, Young et al.~\cite{Young2021_AL_ex2} used active learning to iteratively improve a MLFF that was able to accurately simulate solvents and selected chemical reactions. In a study by Stark et al.~\cite{Stark2024_AL_ex3}, an AL workflow leveraging clustering algorithms was used to model reactive hydrogen dynamics on copper surfaces. Furthermore, Mohanty et al.~\cite{mohanty2023mlff} showed how AL was necessary to augment a dataset for efficiently training MLFFs for polymer dynamics and Kang et al.~\cite{Kang2024_AL_ex9} highlighted how AL was crucial to model strongly anharmonic materials. Numerous other successful AL applications can be found in the literature.~\cite{Zhang2019_AL_ex1,Smith2018_AL_ex2,Zhang2023_AL_ex4, shambhawi2021_AL_ex5,Sivaraman2020_AL_ex6, Kuryla2025_AL_ex7, Erhard2024_AL_ex8,Vandermause2024_AL_ex10,Duschatko2024_AL_ex11,Xie2023_AL_ex12, Vandermause2022_AL_ex13, johansson2022_AL_ex14, Xie2021_AL_ex15, Matsumura2025,gurleksharma2025} 

While AL is always beneficial in the data collection process, the automation degree of the procedure varies broadly. Often, AL is done manually or by users who develop tailored scripts for their specific problems. This situation results in the need for expert knowledge, such as selecting starting geometries, setting uncertainty thresholds, or deciding when to stop sampling. Additionally, employing collections of custom scripts instead of a defined workflow makes the process less accessible to new practitioners and less reproducible by other researchers.
In recent years, the community has started addressing these challenges by offering various automated software solutions. For example, in the DFT codes such as the \textsc{Vienna ab initio simulation package} (VASP)\cite{Kresse1993_VASP1,Kresse1996_VASP2, Kresse1996_VASP3}, \textsc{CASTEP}\cite{Clark2005_castep, Stenczel2023_castep_ml} and the \textsc{Amsterdam Modeling Suite} (AMS)\cite{AMS2025} different  automated AL workflows are implemented. Next to AL methods directly integrated into quantum chemistry codes, there also exist separate software packages offering AL or automated simulation functionalities such as \textsc{FLARE}\cite{Vandermause2020_FLARE}, \textsc{CatFlow}\cite{liu_catflow_2025}, \textsc{Alebrew}\cite{Zaverkin2024}, \textsc{PsiFlow}\cite{Vandenhaute2023_psiflow}, \textsc{ALmoMD}\cite{almomd}, apax\cite{doi:10.1021/acs.jcim.5c01221} or \textsc{PAL}\cite{PAL}.
Although such tools have helped establish MLFFs and AL as a standard tool in molecular simulations, there is a potential for improvements that we address in this work, in particular with respect to the efficiency of configurational space exploration, hardware utilization, support for multi-system sampling and seamless data generation for periodic materials and finite molecular systems.

We present \textsc{aims-PAX}, short for \textit{\textbf{a}b \textbf{i}nitio \textbf{m}olecular \textbf{s}imulation-\textbf{P}arallel \textbf{A}ctive e\textbf{X}ploration}, as a fully automated open-source software package for performing AL, using a parallelized algorithm that enables efficient resource management. The current implementation is integrated with the \textsc{FHI-aims}~\cite{aims-roadmap-2025} program for DFT calculations and the \textsc{MACE}~\cite{Batatia2022mace, Batatia2025design} architecture as an MLFF model.
However, while the software is developed primarily for working in tandem with the codes and models named above, the algorithm itself is agnostic to both the MLFF architecture and the choice of DFT code. 

Importantly, we want to highlight defining features of \textsc{aims-PAX} that emphasize its versatility and uniqueness:

\begin{enumerate}
    \item Leverage of GP-MLFFs for data acquisition
    \item Multi-system sampling for transferable MLFFs
    \item Flexible combination of \textit{ab initio} levels of theory
    \item Seamless handling of molecular and materials systems
    \item Efficient CPU/GPU workload management
    \item Support for state-of-the-art neural network based MLFFs
\end{enumerate}

We demonstrate the capabilities of \textsc{aims-PAX} herein on a flexible peptide, where it autonomously generates accurate and stable MLFFs using two orders of magnitude fewer DFT calculations than traditional workflows.\cite{tea_1, tea_2} Beyond individual systems, \textsc{aims-PAX} concurrently samples multiple molecules, autonomously identifying the most informative configurations to build a single, transferable MLFF that generalizes across chemical space. Integrated seamlessly with \textsc{FHI-aims}, it unifies gas-phase, bulk, and solvated regimes—capturing paracetamol in vacuum, microsolvated by water, and explicit solvent within one model. Finally, large-scale tests on a bulk perovskite demonstrate its exceptional scalability and computational efficiency, establishing \textsc{aims-PAX} as a universal framework for automated, data-driven force-field generation.


\section{Results}
\label{sec:methods}

\begin{figure*}
 \centering
 \captionsetup{justification=centering}
 \includegraphics[scale=0.9]{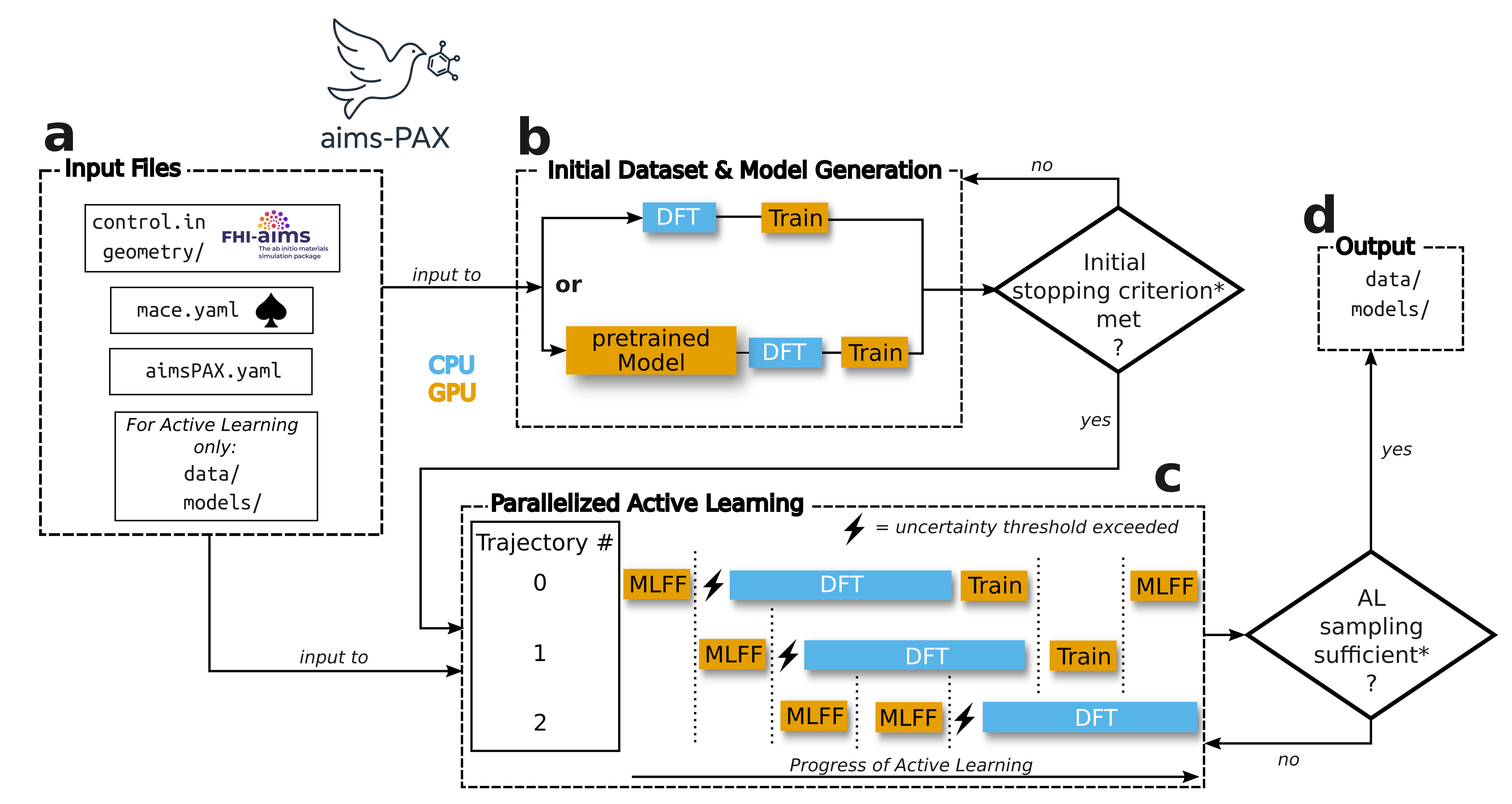}
 \caption{
\textbf{Overview of the \textsc{aims-PAX} workflow:}
(a) Required input files: The first file (\code{control.in}) follows FHI-aims conventions\cite{Blum2009} and contains the DFT settings. It is also possible to use different DFT settings per trajectory. The system's geometry, or initial geometries can either be inside a folder (\code{geometry/}) or, in the case of a single geometry, in a file (\code{geometry.in}). The file (\code{mace.yaml}) contains MACE\cite{Batatia2022mace, Batatia2025design} model hyperparameters and the fourth (\code{aims_PAX.yaml}) is an \textsc{aims-PAX}-specific file containing the IDG and AL settings. For the AL workflow, folders containing the initial datasets (\code{data/}) and models (\code{models/}) are required. (b) Initial dataset generation (IDG): Geometries are sampled using either DFT or a GP model, with DFT providing labels in both cases. Sampling continues until a (* user specified) criterion is met. (c) Parallelized active learning: The AL workflow requires input files, existing data, and models, which can be provided by the IDG procedure. Sampling occurs over multiple trajectories, triggering DFT calculations when an uncertainty threshold is exceeded. GPU-based ML tasks (orange) and CPU-based DFT tasks (blue) can run in parallel. AL is continued until a (* user specified) stopping condition is met. (d) Output: Models and collected data produced during AL (and IDG).  
 }
 \label{fig:al_parallel}
\end{figure*}

\subsection{Initial Dataset and Model Generation}
\label{sec:initial_ds}

A starting point for an AL procedure involves generating an initial ensemble of MLFFs, or a single MLFF, that simultaneously predicts the potential energy surface (PES) and associated uncertainties, capable of producing stable molecular dynamics within a limited region of the PES. We want to emphasize that this part of the workflow is not \textit{active} in the sense that the model does not choose which points to include in the training. At this stage, the model is not yet sufficiently reliable to guide this selection process. Thus, data is generated using a sampling strategy, such as molecular dynamics (MD).

Such an initial dataset generation (IDG) can, for example, be established using one of two approaches:

\begin{enumerate}
    \item Short \textit{ab initio} simulations can be run to generate molecular configurations  along with their respective energies, forces etc.
    \item A GP-MLFF can be used to produce physically plausible system geometries. These geometries are then recomputed using a reference \textit{ab initio} method.
\end{enumerate}

The second approach is generally preferable, as it helps decorrelate the geometries, making the IDG significantly more computationally efficient.
Both initialization strategies are implemented in \textsc{aims-PAX}, see Fig.~\ref{fig:al_parallel}b. 
We also couple the IDG with \textsc{Parsl}\cite{babuji19parsl}, similar to what is done in \textsc{Psiflow}\cite{Vandenhaute2023_psiflow}. This enables users to perform DFT calculations on sampled geometries across multiple nodes in parallel. Importantly, the GP model does not need to provide accurate energies or forces; it acts solely as a geometry generator in combination with MD simulations.
Currently, the implementation includes the MACE-MP \cite{batatia2024foundationmodelatomisticmaterials} and SO3LR~\cite{Kabylda2025_so3lr} GP models, with additional models to be incorporated in the future.

Once the dataset reaches a user-defined threshold in size, it is split into several equally sized subsets. These subsets are randomly selected from the full dataset, with one subset being assigned to each MLFF ensemble member. This ensures that each model is trained and validated on slightly different data. In addition to varying the training data, we introduce further diversity between ensemble members by using different random seeds for initializing model weights. Each MLFF is then trained on its assigned subset for a user-specified number of epochs.
Optionally, this procedure of generating data and training can be repeated multiple times. More precisely, after training, new structures are sampled and subsequently labeled, which is followed by more training steps. More details on this approach can be found in Section~S5A.

During these cycles, the MLFFs are trained
without reinitializing their weights. Instead, the existing weights are reused at every training step, and models are trained on their entire datasets to prevent catastrophic forgetting\cite{French1999_forgetting1, McCloskey1989_forgetting2, McClelland1995_forgetting3, Ratcliff1990_forgetting4, Kumaran2016_forgetting5}. This continual learning (CL) approach \cite{Kirkpatrick2017_CL1, aljundi2019_CL2, Parisi2019_CL3} enables models to improve iteratively without retraining from scratch each time, reducing the number of required training steps.
Crucially, at this stage, models are not trained to full convergence; instead, training is deliberately limited to a small number of epochs. The described early-stopping strategy avoids overfitting the models on the initial datasets and hinders them from getting stuck in local minima. This would make updating the models with new data during the AL significantly more difficult without reinitializing their weights.

The IDG is repeated until a user-defined stopping criterion is met. Possible stopping criteria include a maximum number of training epochs, a predefined training set size, or a target performance (e.g., force mean absolute error, MAE) on the validation set. The latter can be aligned with the overall AL workflow termination condition. For instance, the user may specify a target force MAE that should be achieved on the validation dataset by the end of the AL process. A scaling factor can be applied to this target MAE to define the stopping criterion for MLFF ensemble pretraining. At this stage, the goal is not to develop highly accurate MLFFs or exhaustive datasets but to obtain a robust MLFF ensemble capable of generating stable dynamics within an initial region of the PES, from which the main AL workflow can begin sampling the broader PES landscape.

\subsection{Parallelized Active Learning}
\label{sec:active_learning}

The AL phase involves sampling the configurational space of the target system using a pre-trained ensemble of MLFFs, which are employed for both sampling and uncertainty quantification. The latter is used together with a threshold that determines when a sampled structure is supposed to be labeled \textit{via} a DFT calculation.

In the case of \textsc{aims-PAX}, each time the threshold is crossed and the DFT calculation has been performed, the new data is added to the training (or validation) set of all MLFFs. These are then updated in a CL scheme using a user-specified, ideally low, number of epochs similarly to the one employed in the IDG. For more details on the exact training strategy we refer to Section~S5A. 
While the algorithm proposed herein is, in principle, agnostic to the choice of uncertainty quantification method, we employ the \textit{query by committee} (QBC)\cite{Seung1992_QBC, schran2020committee, nong2012Committee} approach due to its conceptual simplicity and widespread adoption. The integration of alternative uncertainty estimation techniques into our framework is straightforward and will be explored in future work. \\
\indent In the QBC approach, an ensemble of independently trained ML models is used to produce a distribution of predicted outputs during inference. As described previously, diversity among ensemble members arises from differences in initial weight initialization seeds and distinct initial training datasets. The variance within the ensemble predictions serves as a way to quantify a model's uncertainty. Specifically, we quantify uncertainty based on the variance of atomic force predictions, using the maximum per-atom force variance across the system, as defined in Eq.~\ref{eq:uncertainty_measure}\cite{Zhang2019_AL_ex1},

\begin{equation}
\label{eq:uncertainty_measure}
\delta_n = \underset{i}{\max}\sqrt{\frac{1}{3M}\sum^M_{j=1}\sum_{k\in{x,y,z}}\left(F_{nijk}-\Bar{F}_{nik}\right)^2},
\end{equation}
where $\delta_n$ denotes the uncertainty associated with geometry $n$. The maximum is computed over all atoms $i$ in the system. The ensemble consists of $M$ models indexed by $j$, and the summation over $k$ spans the three spatial components $x$, $y$, and $z$. The term $F_{nijk}$ represents the $k$-th Cartesian component of the force on atom $i$ in system $n$ predicted by model $j$, while $\Bar{F}_{nik}$ denotes the ensemble-averaged force component on atom $i$ in direction $k$.

For setting the uncertainty threshold, we adopt an approach analogous to the one implemented in \textsc{VASP}\cite{Kresse1993_VASP1, Kresse1996_VASP2, Kresse1996_VASP3}, where a scaled moving average of the uncertainties is used in place of a fixed threshold. Specifically, the threshold at iteration \( t \), denoted by \( \delta_t \), is computed using Eq.~\ref{eq:threshold},

\begin{equation}
    \label{eq:threshold}
    \delta_t = \frac{1+c_x}{N} \sum_{n=1}^{N} \delta_n.
\end{equation}

Here, \( N \) represents the number of past uncertainty values included in the moving average, for which we follow definitions introduced in the \textsc{VASP} code and use a default window size of 400. The scaling factor \( c_x \) allows the threshold to be adjusted: values \( c_x < 0 \) tighten the threshold, while \( c_x > 0 \) relax it. In our implementation, the default value is \( c_x = 0 \). We also include the option to freeze the threshold after a user-specified training set size.

The primary advantage of this adaptive-threshold approach is that it eliminates the need for a fixed, user-defined uncertainty cutoff, which can vary between systems.\cite{Kulichenko2023} Since the moving average naturally decreases over time, some configurations will always exceed the threshold. As a result, the sampling frequency depends on the value of \( c_x \): if set too high, very few points may be sampled; if too low, the method may oversample.  Based on our experience and also reported for the MLFF training in the \textsc{VASP} code, values of \(c_x \in \left[-0.1, \,0.1\right]\)  serve as practical starting points.

To improve the efficiency and robustness of the active sampling, we adopt a multi-trajectory approach that has also been successfully applied in similar frameworks.\cite{liu_catflow_2025, PAL, Matsumura2025} Herein multiple ML-driven simulations are executed in parallel, see Fig.~\ref{fig:al_parallel}c. These trajectories may differ in their sampling strategies, utilizing various thermostats, barostats, external conditions, or simulation schemes. Importantly, different trajectories can also simulate different systems. We point out that the uncertainty threshold as defined in Eq.~\ref{eq:threshold} is shared across all of these trajectories.

A key advantage of multi-trajectory sampling is its ability to decouple the generation of new configurations from the evaluation of high-uncertainty states. While MLFFs generate new candidate geometries, DFT calculations are performed in parallel on selected high-uncertainty configurations to enrich the reference dataset. These calculations are done using \textsc{FHI-aims}\cite{aims-roadmap-2025} compiled as a library and interfaced through the  \textsc{Atomic Simulation Interface}\cite{Stishenko2023}. The latter allows to run an instance of \textsc{FHI-aims} continuously, meaning that the DFT code does not have to be reinitialized before every calculation. This can remove significant overhead, which is especially valuable when handling smaller systems. 

As mentioned earlier the training is done using a CL scheme, similar to the one used during pre-training, which allows MLFFs to be incrementally updated during sampling. Together with the parallel DFT calculations, this strategy also optimizes utilization of available computational resources (CPUs and GPUs), thereby enhancing the overall efficiency and throughput of the AL workflow.

Similarly to the IDG, \textsc{Parsl}\cite{babuji19parsl} can also be used here, allowing to distribute DFT calculations across multiple nodes in parallel. Additionally, the number of DFT workers can be adapted dynamically up to a user-defined maximum. Through the flexible allocation of resources, we ensure that no worker is idle or overloaded. This is particularly useful in AL as the demand for new data can change during the procedure. For example, there can be long sequences where an MLFF trajectory is certain about all encountered geometries or intervals where all trajectories lead to high uncertainty regions. Due to these crucial computational benefits this approach has been designated as the default procedure in \textsc{aims-PAX}.

As with the IDG, the AL workflow proceeds until a user-defined stopping criterion is met. This may be based on the total training set size, performance on the validation set (e.g., force MAE), number of training epochs, or total number of MD steps. Once the stopping criterion is met, either the entire ensemble or only the best-performing ML model, selected based on validation error, is further trained to converge on the whole training set.

\subsection{The phase-space of a peptide: Ac-F-A5-K}
\label{sec:application_pepitde}

\begin{figure*}
 \centering
 \captionsetup{justification=centering}
 \includegraphics[scale=0.9]{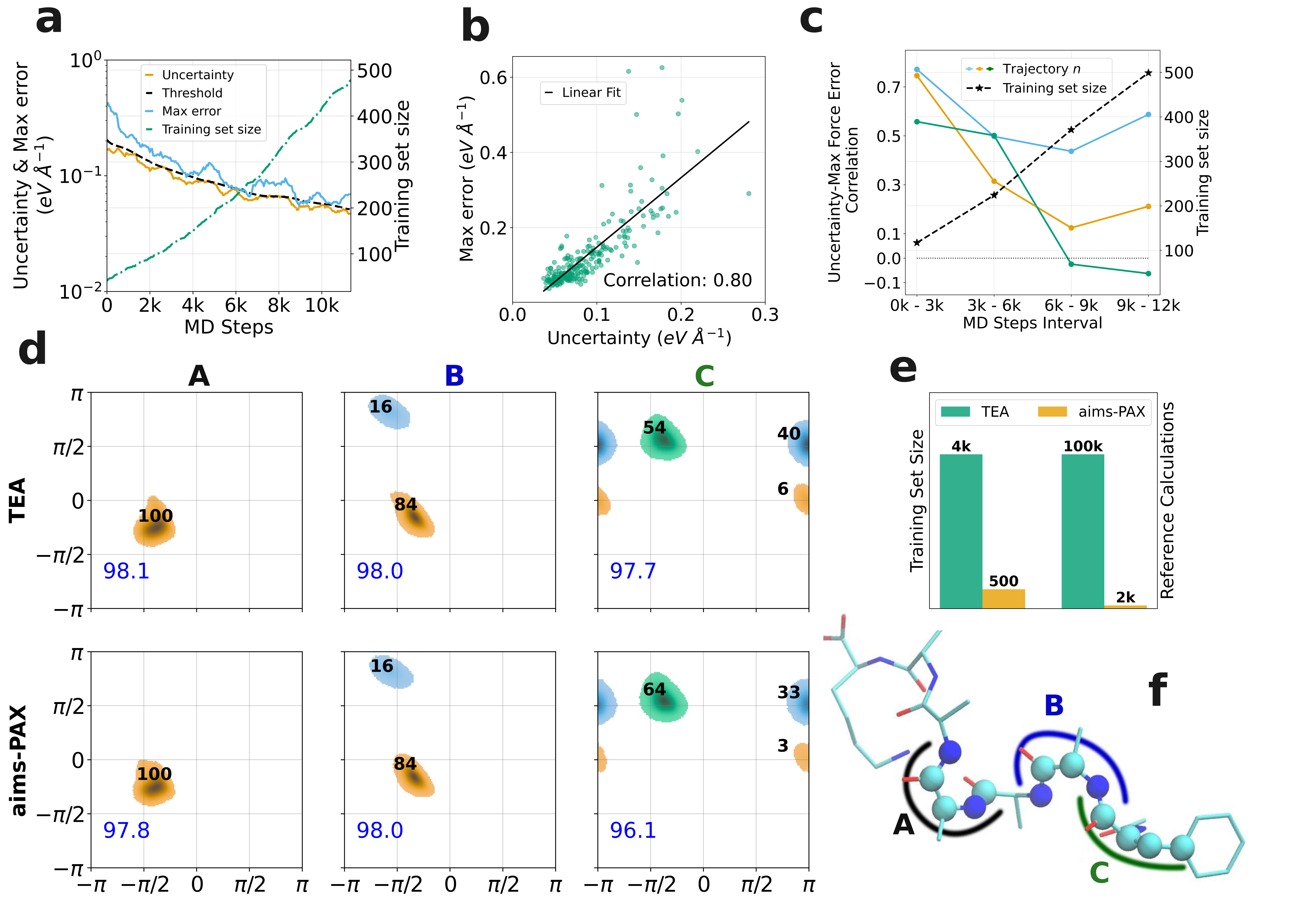}
 \caption{
 \textbf{\textsc{aims-PAX} applied to the peptide Ac-F-A5-K:} (a) Model uncertainty, actual maximum force error, uncertainty threshold and training set size as a function of MD steps throughout the AL procedure. (b) Actual maximum force error vs. model uncertainty with Pearson correlation coefficient over the whole AL workflow. A linear fit is shown as a guide to the eye. (c) Pearson correlation coefficient and training set size over multiple segments of the AL workflow for $n=3$ trajectories that were used for sampling. (d) Ramachandran plot for selected dihedral angles (see f) acquired with a model used in the TEA challenge\cite{tea_1, tea_2} (left) and ours, acquired using \textsc{aims-PAX} (right). Relative populations of highlighted clusters are given in bold font (black) and the blue number in the bottom left corner of each plot indicates the percentage of configurations from the MD trajectories assigned to a cluster.\cite{tea_1,tea_2} (e) Number of geometries in the training set (left axis and bars) and number of required reference calculations for the dataset creation (right axis and bars) using a manual approach (as done in the TEA challenge,\cite{tea_1,tea_2} green) and \textsc{aims-PAX} (orange)  (f) Structure of Ac-F-A5-K including highlighting of relevant dihedral angles A,B and C. 
 }
 \label{fig:peptide_results}
\end{figure*}

To demonstrate the performance of \textsc{aims-PAX}, we first use it for a challenging, isolated system: a N-acetylphenylalanyl-pentaalanyl-lysine (Ac-F-A5-K) peptide. This system was selected because of its complexity and relevance to typical MLFF applications in biochemistry. Our results demonstrate that the proposed AL framework reduces the number of required reference evaluations by up to three orders of magnitude and substantially minimizing the necessary human effort.

This peptide exhibits multiple local minima explored during MD simulations under ambient conditions, which typically require numerous costly reference calculations when using conventional, non-AL methods. To assess the reliability of uncertainty estimates within our AL workflow, we trained an ensemble of models for the Ac-F-A5-K peptide using the \textsc{aims-PAX} framework, based on three parallel MD trajectories, as detailed in Section~\ref{sec:comp_details}. During the AL process, we also perform DFT reference calculations every 50 MD steps independently from the uncertainty selection criterion. Additionally, the actual prediction error and model uncertainty were evaluated at these  points. Using this data we analyze the behavior of the uncertainty measure throughout the AL procedure without a bias towards high uncertainty states.\\
\indent Fig.~\ref{fig:peptide_results}a shows the evolution of prediction error, uncertainty threshold, and training set size over the course of an AL run for each trajectory. All three trajectories display consistent behavior: model uncertainty, the uncertainty threshold, and prediction error all decrease systematically as AL progresses. Notably, the temporal profiles of uncertainty and error follow similar trends, indicating a positive correlation between these quantities.
To quantify this observation, we plot the uncertainty against the maximum atomic force error in Fig.~\ref{fig:peptide_results}b, along with a linear regression fit, and compute the Pearson correlation coefficient. Across all trajectories, we observe a clear positive correlation between uncertainty and error, with only a limited number of outliers. This positive correlation is crucial, confirming that ensemble uncertainty can serve as an effective proxy for prediction error. Consequently, the AL algorithm selectively targets challenging configurations for high-fidelity DFT calculations while avoiding redundant sampling of trivial structures. Importantly, a perfect agreement between uncertainty and error is not required for practical applications; some errors may be missed in the early stages but captured at later AL steps as more diverse geometries are encountered.

Despite the widespread use of the QBC strategy for MLFFs, concerns have been raised regarding its reliability.\cite{scheffler2023_uncertainty} Also, we have chosen a relatively small ensemble size of 4 and it has been reported that small ensembles result in biased estimators of uncertainties and other properties~\cite{imbalzano2021uncertainty}. However, it is not unusual, to use only a few ensemble members for AL in MLFFs.\cite{Zhang2019_AL_ex1, Stark2024_AL_ex3, Kuryla2025_AL_ex7,Erhard2024_AL_ex8,Kang2024_AL_ex9} This is often done to reduce the computational expense of an AL procedure as increasing the number of ensemble members means more ML models have to be trained and evaluated.
Indeed, as shown above, the uncertainty measure that we obtain is already a good approximation for the real error with four members. Thus, there is no need to incur greater computational cost by using more MLFF models in the ensemble.

To further probe the reliability of our approach, we analyze the evolution of the Pearson correlation coefficient between uncertainty and error throughout the AL process, see Fig.~\ref{fig:peptide_results}c. In the initial 3k MD steps, the correlation exceeds 0.5 for all three trajectories. However, the correlation declines from 3k to 9k MD steps, even turning negative for the third trajectory (green).
This degradation in uncertainty quality may be attributed to increasing overlap among the training sets of individual ensemble members as the AL progresses. As the models are exposed to similar data, they tend to converge on the same underlying potential energy surface, thereby reducing ensemble diversity. Nonetheless, the use of multiple trajectories helps alleviate this issue. A  significant correlation for even a single trajectory can drive effective data acquisition, ensuring the continued efficacy of the overall AL scheme.

Another important aspect of the proposed AL workflow is the influence of multiple concurrent trajectories on the sampling process. To evaluate how model accuracy depends on the number of parallel trajectories used during AL, we conducted multiple \textsc{aims-PAX} runs with varying numbers of concurrent MD simulations. We performed three independent AL runs with different random seeds per setup for statistical reliability. For subsequent tests, we selected the best-performing model (based on validation set accuracy) from each of these three independent runs, resulting in three models per setup.

The test sets were generated by performing $1$~ns of NVT MD at 300, 500, and 700~K using the MACE-OFF (small) potential\cite{Kovcs2025_maceoff}. Representative structures were selected from these trajectories using farthest point sampling (FPS) based on ML-derived descriptors. Reference energies and forces were then computed at the chosen level of theory. Additional computational details are provided in Section~\ref{sec:comp_details} and a comprehensive account of the results is given in Section~S1.

We could not observe a dependence of accuracy on the number of sampling trajectories used in \textsc{aims-PAX}. The comparable model accuracies at given temperatures across all setups indicate that the AL-generated datasets are of similar quality. Notably, the total number of MD steps required to gather the training data remained approximately constant across all settings. For instance, a run with a single trajectory required an average of $\sim$68k MD steps to collect 1k structures (500 for training and 500 for validation). In contrast, setups using 8 and 32 trajectories converged after only $\sim$9k and $\sim$2k MD steps per trajectory, respectively. These findings and the above-mentioned improvement in the uncertainty measure robustness for the multi-trajectory approach suggest that increasing the number of trajectories improves sampling efficiency without compromising data quality. 

To investigate whether the use of CL during AL influences the performance of the resulting MLFFs, we retrained new models from scratch using the datasets acquired throughout the AL process. These models were evaluated using the same protocol described above for multiple trajectories. No deterioration in performance was observed for the models trained with CL compared to those trained from scratch. Detailed results are presented in Section~S2. The results confirm that the continuous learning paradigm offers a more computationally efficient alternative to repeated retraining without compromising the accuracy or robustness of the final MLFF models.

An essential requirement for MLFFs is the stability of the resulting MD simulations.~\cite{fu2023forces}. We performed four 1~ns-long NVT MD runs with each of the three models at 300, 500, and 700~K. This resulted in a total of 12 MD runs per number of trajectories used in the AL procedure and temperature. We define a simulation as stable if no covalent bond in the system exceeds 2~\r{A}, a condition that is not expected to be violated at the temperatures considered. For more details on the MD themselves, see Section~\ref{sec:comp_details} and for a thorough report on the number of stable runs of each run, see Section~S1. 

Overall, no clear trend emerges linking the stability of the MLFFs to the number of trajectories used during AL. This suggests that, for the current AL setup, model robustness in MD simulations is not significantly affected by the number of concurrent sampling trajectories. This behavior may be attributed to all trajectories using the same sampling protocol, potentially limiting exploration diversity. Future work will explore diverse sampling strategies across trajectories during AL to improve coverage of the potential energy surface and enhance model robustness under elevated temperatures and extreme simulation conditions. 

Finally, the most reliable validation of an MLFF model lies in evaluating its performance in realistic application scenarios. Here, we assess the model’s ability to reproduce the Ramachandran plots from molecular dynamics simulations conducted under ambient conditions. The procedure follows that of the TEA 2023 Challenge benchmark~\cite{tea_2}.

We take the Ramachandran plots produced by the MACE model trained on the complete dataset in Ref.~\onlinecite{tea_2} as a reference. We recomputed the structures sampled by the \textsc{aims-PAX} workflow, using three parallel AL trajectories, at the same level of theory employed in the TEA 2023 Challenge (PBE0+MBD-NL/\code{intermediate}). A new MACE model was then trained using the same architecture and hyperparameters as the reference study. For further details, see Section~\ref{sec:comp_details}. This recomputation was necessary because, during the AL phase, we employed a smaller MACE model trained on PBE+MBDNL/\code{light} to reduce computational costs. Such sampling-by-proxy strategies are commonly used in MLFF development~\cite{mlff_review}, and we demonstrate here how \textsc{aims-PAX} can efficiently generate high-quality, diverse datasets with minimal DFT overhead.\\
\indent The retrained model was used to perform 12 independent $1$~ns NVT MD simulations at 300~K, each initialized from a different starting geometry, following the protocol of the TEA 2023 Challenge. The resulting trajectories were analyzed by extracting dihedral angle distributions, which were then clustered following the methodology from Ref.~\onlinecite{tea_2}. The Ramachandran plots obtained from our model and the TEA reference model are shown in Fig.~\ref{fig:peptide_results}d.
Both this work's and the reference MACE models yield nearly identical cluster structures and populations for dihedral angles A and B, which correspond to the peptide backbone. Specifically, for angle A, a single dominant cluster is located at approximately ($-\pi/4$, $\pi/2$) (blue), with a relative population of 100\%. For angle B, two clusters appear in both models: one at ($-\pi/4$, $\pi/2$) (blue) and another at ($-\pi/2$, $\pi$) (green), with relative populations of 84\% and 16\%, respectively.\\
\indent Minor differences are observed only in the dihedral angle C, which pertains to the peptide tail. Both models identify three clusters at similar angular positions, but relative populations differ slightly. For the blue cluster at ($-\pi/2$, $\pi/2$), our model predicts a population of 64\%, compared to 60\% in the reference model. The orange cluster at ($\pm\pi$, $0$) appears with a population of 7\% in our model and 3\% in the reference. The green cluster at ($\pm\pi$, $\pi/2$) is equally represented in both cases, with a population of 33\%. The observed differences between MD results can likely be attributed to limited sampling statistics, as capturing slow conformational changes at the peptide tails may require simulations significantly longer than 12~ns.\\
\indent A crucial advantage of the proposed AL workflow is that our model was trained on only 500 reference structures, requiring a total of just 2,000 DFT calculations—including those performed during the AL process and the subsequent recomputation at a higher level of theory (see Fig.~\ref{fig:peptide_results}e). In comparison, the reference model in Ref.~\onlinecite{tea_2} was trained on 4,000 structures generated from 100,000 DFT calculations, a process that also involved several months of manual effort. These results highlight the efficiency and scalability of the \textsc{aims-PAX} framework for the automated generation of high-quality training datasets. In particular, we demonstrate a reduced number of DFT evaluations by up to three orders of magnitude, achieved with minimal human intervention, while obtaining a final MACE model that delivers comparable predictive performance. Future developments, including the implementation of more reliable uncertainty quantification methods and diverse sampling techniques, are expected to strengthen further the advantages of the proposed automated AL workflow over traditional dataset generation and MLFF training approaches.

\subsection{MD17: Sampling Chemical Space of Small Molecules}
\label{sec:multi-sys sampling}

  \begin{figure*}
 \centering
 \captionsetup{justification=centering}
 \includegraphics[scale=0.9]{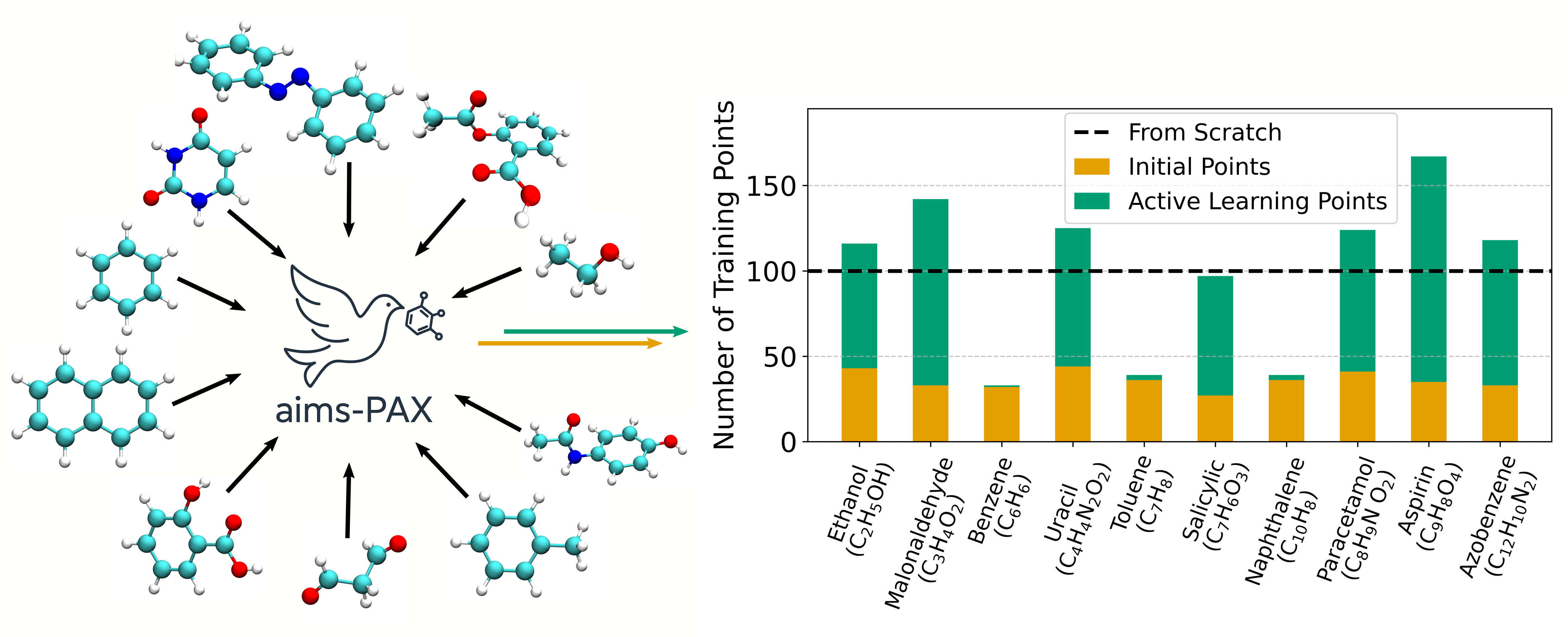}
 \caption{
     \textbf{Creation of a transferable MLFF \textit{via} \textsc{aims-PAX} through astute sampling:}
     Number of data points in the training sets of the MLFF acquired using \textsc{aims-PAX}. The data is split up in points attained through the initial dataset generation (yellow) and the active learning itself (green). The model trained from scratch through a manual data curation approach uses 100 points for each chemical species (black dashed line).}
 \label{fig:md17_results}
  \end{figure*}

The multi-trajectory sampling approach in \textsc{aims-PAX} can also be used to generate data for different chemical species at the same time. To illustrate this, we have chosen the molecules from the MD17 dataset\cite{chmiela2017}. We ran \textsc{aims-PAX} where each molecule is assigned to a different trajectory and an ensemble of models is trained on all systems during AL. The best performing MLFF of this ensemble is then used for evaluation.

For comparison, we train a separate MLFF from scratch using geometries from the first half of each trajectory in the MD17 dataset. The exact data selection process is described in Section~\ref{sec:comp_details}. This analysis is done to compare an MLFF and its dataset acquired through \textsc{aims-PAX} with the one obtained by a more manual and ``traditional'' approach. It is worth noting that, similarly to the comparison done in Section~\ref{sec:application_pepitde}, while the model trained from scratch is trained on the same amount of data as the model obtained with \textsc{aims-PAX}, \textit{i.e.} 1,000 training points, the acquisition of this data required roughly two or even three orders of magnitude more DFT calculations. In the case of malonaldehyde in MD17, for example, AIMD simulations containing almost 1 million steps were performed.

We have chosen the systems in MD17, as they offer various PES complexities. Molecules such as benzene or toluene are rigid and highly symmetric, making them easy to learn for MLFFs. In contrast, the PES of flexible molecules such as aspirin or azobenzene is more challenging to reproduce. However, it is not always known \textit{a priori} which systems an MLFF will struggle with and by how much. Consequently, it is also unclear which geometries should be added to a training set and in what quantity. Ideally, during AL, the uncertainty measure should automatically select challenging systems to avoid this problem.

The results of this study are illustrated in Fig.~\ref{fig:md17_results}. In the right panel of the figure, we show the number of geometries for each molecule in the training data set for the model acquired through \textsc{aims PAX} and the training from scratch. The latter has 100 points for all species (black, dashed line), which have been obtained by randomly sampling from subsets of the respective datasets in MD17 (for more details see Section~\ref{sec:comp_details}). In contrast, the model from \textsc{aims-PAX} has a varying number of points per molecule. We also make a distinction between the points from the IDG (see Section~\ref{sec:initial_ds} for more details) depicted in yellow and the points that were acquired during the actual AL. The latter are shown as green bars stacked on top of the yellow ones.
During AL, most points were sampled for aspirin (132 new geometries), malonaldehyde (109), and azobenzene (85), while the smallest amount of points were added for toluene, naphthalene (both 3), and benzene (1). These values align with our expectations formulated above. More data was collected for challenging and flexible molecules compared with those for simpler and more rigid ones.

Although the number of training points supports our assumption about the challenge that each molecule poses for the MLFF, we extended our analysis by investigating the accuracy of the acquired models. These results are shown in Fig.~S1 of Section~S3 and we summarize the key finding here.
In essence, it was observable that the lowest errors were achieved for benzene and naphthalene, respectively. This supports the notion that \textsc{aims-PAX} mostly picks challenging systems for labeling using DFT, saving computational effort from being expended on easily learnable molecules. The largest errors were obtained with aspirin and malonaldehyde. To address these challenging systems, \textsc{aims-PAX} automatically sampled substantially more configurations for these molecules than for simpler ones. Consequently, the resulting MLFF achieved lower errors compared to the traditionally created MLFF. This demonstrates further that \textsc{aims-PAX} adapts naturally to differences in molecular complexity without manual intervention, yielding models that perform better on difficult systems than those trained from manually curated datasets.

We also point out that the model obtained with \textsc{aims-PAX} was capable of running stable 1 ns long MD simulations at 300~K for any of the molecules in MD17. Overall, this underscores the suitability of \textsc{aims-PAX} for efficiently generating balanced datasets across multiple systems with potential applications in curating or even training GP MLFFs from scratch.

\subsection{Paracetamol in Water: Simultaneous Sampling of Periodic and Non-Periodic Systems}

\begin{figure*}
 \centering
 \captionsetup{justification=centering}
 \includegraphics[scale=0.9]{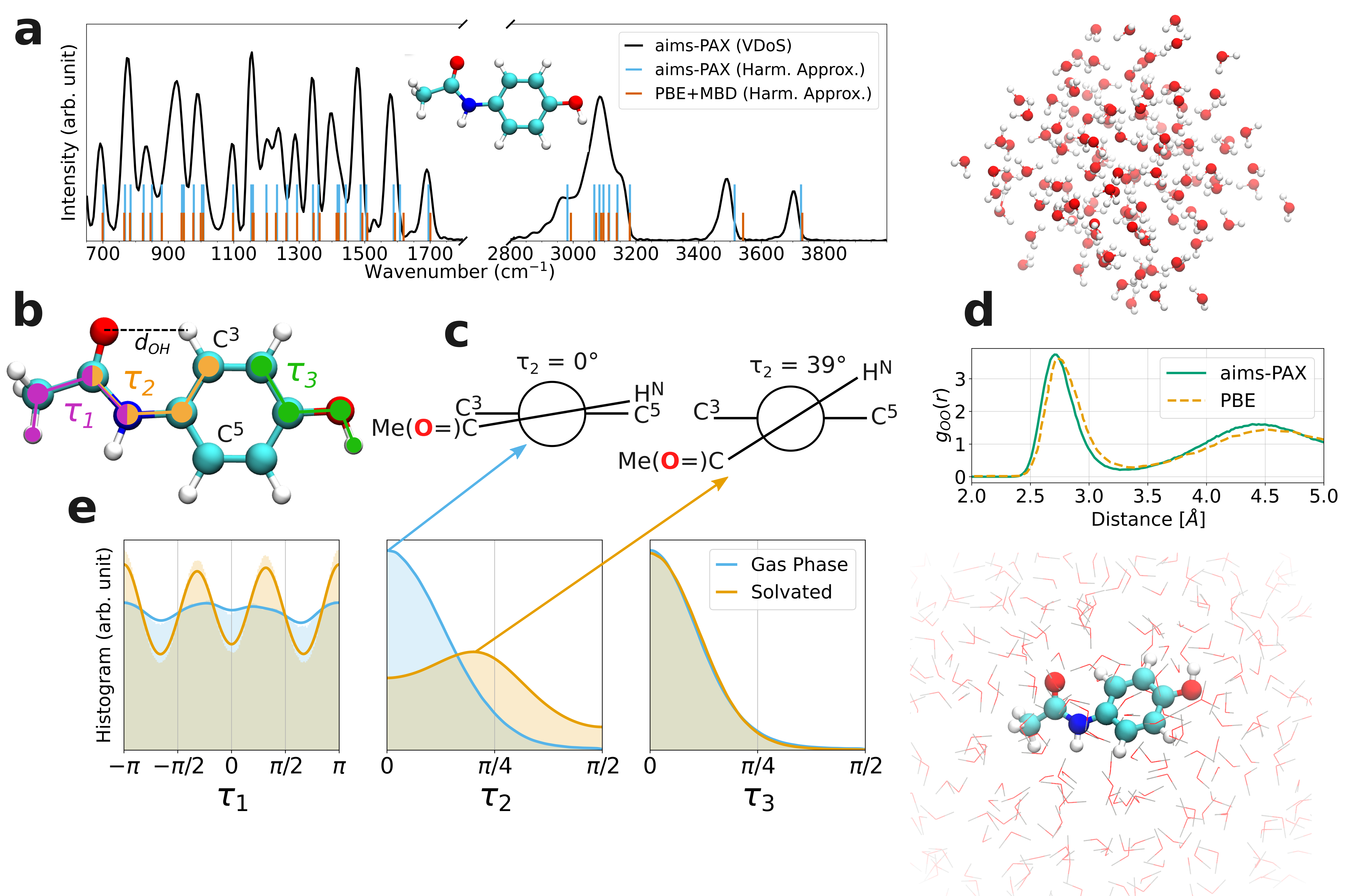}
 \caption{
    \textbf{\textsc{aims-PAX} used for creating a model capable of modeling explicit solvation:} a) Vibrational density of states (VDoS) for paracetamol in the gas phase at 300~K acquired from the velocity autocorrelation function using the MLFF (solid black line) compared to the vibrational frequencies acquired within the harmonic approximation using said MLFF (tall blue vertical lines) and DFT (short deep orange lines). b) Depiction of paracetamol with highlighted atoms that define the three dihedral angles analyzed in this work ($\tau_1$ in magenta, $\tau_2$ in orange, and $\tau_3$ in green) as well as the definition of $d_{OH}$ and marking of carbons three and five used in c) of the same figure. c) Newman projection\cite{newman_proj} along $\tau_2$ of paracetamol for the cases $\tau_2=0^{\circ}$ and $\tau_2=39^{\circ}$ corresponding to the maxima in e). d) Snapshot of an MD trajectory of bulk water and the oxygen-oxygen radial distribution function obtained from simulations run by the MLFF acquired from \textsc{aims-PAX} and AIMD using PBE\cite{water_rdf_pbe}. e) Histogram and associated kernel density estimation of dihedral angles $\tau_1$, $\tau_2$, and $\tau_3$ from simulations of paracetamol in gas phase and explicit water. Simulations were run using the MLFF acquired through \textsc{aims-PAX}, and a snapshot of the simulation with solvent is depicted.
    }
 \label{fig:solv_results}
\end{figure*}

To further emphasize the multi-system sampling capability and leverage the strengths of \textsc{FHI-aims}, we employ \textsc{aims-PAX} to sample trajectories from both periodic and non-periodic structures simultaneously. The use of \textsc{FHI-aims} is essential here, as it enables seamless and efficient DFT calculations across bulk and isolated systems within a unified framework. This capability is particularly valuable for generating consistent datasets and machine-learning models that capture the physics of realistic, extended materials.

As an example, we construct a model capable of performing simulations of an explicitly solvated paracetamol molecule, an isolated paracetamol molecule in the gas phase, and bulk water. For this we run \textsc{aims-PAX} with multiple trajectories consisting of paracetamol in the gas phase, the same molecule surrounded by a cluster of 90 water molecules, and bulk water containing 64 water molecules with periodic boundary conditions. Thus, data is acquired that enables the model to learn the intramolecular interactions of the solute, interactions of the solute with the solvent, and the solvent in itself. The exact settings are given in Section \ref{sec:comp_details}.

The final training set consists of 85 bulk water structures, 185 structures of the isolated paracetamol molecule, and 730 solvated paracetamol structures. As discussed in the previous section, \textsc{aims-PAX} naturally samples more challenging and informative data points. The fact that bulk water was sampled with the fewest amount of points is explained by the fact that a single instance of a periodic bulk structure contains more information on the interactions governing the system than an isolated system would. A single frame of bulk water contains many examples of inter- and intramolecular interactions for the same system. In contrast, paracetamol surrounded by a water cluster is a challenging system that includes interactions of the solute and solvent. Furthermore, while paracetamol in the gas phase is a comparatively simple system, the model has to learn its differing behavior in the gas phase and when solvated, resulting in the second largest fraction of the final training data.

In order to test the resulting model, we elucidate its capability of handling paracetamol in the gas phase, simulating bulk water and explicitly solvated paracetamol. We want to stress that the goal here is not to generate a model that could, e.g., simulate water in a highly realistic and accurate manner. Our tests are designed to demonstrate that \textsc{aims-PAX} enables the efficient construction of a unified model capable of treating both periodic and aperiodic systems within a single framework, yielding stable dynamics that accurately reflect the reference level of theory and exhibit physically consistent behavior.

For testing the model on gas-phase paracetamol, we generated the vibrational density of states (VDoS) by performing multiple independent MD simulations. The VDoS was obtained from the velocity autocorrelation function followed by a Fourier transform. Additionally, vibrational frequencies were computed within the harmonic approximation using both the ML model and DFT reference calculations for comparison. A detailed description of the employed methods is provided in Section~\ref{sec:comp_details}, and the corresponding results are summarized in Fig.~\ref{fig:solv_results}a.

First, it can be seen that the positions of the vibrational frequencies within the harmonic approximation acquired with DFT and the MLFF from \textsc{aims-PAX} coincide for nearly all instances. Notable exceptions are around wavenumbers of 1600~$\mathrm{cm}^{-1}$, 2990~$\mathrm{cm}^{-1}$, and between 3500 and 3550~$\mathrm{cm}^{-1}$. The discrepancy for the latter is the largest. This spectral region is associated with vibrations of the hydroxyl group. The model shows limited ability to reproduce the exact gas-phase behavior when trained on a combined gas-phase and water-cluster dataset, a shortcoming that stems from architectural constraints rather than the active learning strategy itself.

The peaks of the anharmonic VDoS align in general with the vibrational frequencies acquired using the harmonic approximation. Around 3500~$\mathrm{cm}^{-1}$ and 3700~$\mathrm{cm}^{-1}$, a shift to lower frequencies can be observed. A broad signal between 2900~$\mathrm{cm}^{-1}$ and 3200~$\mathrm{cm}^{-1}$ is observable. This region corresponds to vibrations for the NH bond in the amide, stretching of the aromatic bonds, and $\mathrm{sp}^3$ C-H bonds. The signal for the stretching of the CO double bond is visible around 1600~$\mathrm{cm}^{-1}$.\cite{spectra} In total, these results show that the model is capable of correctly reproducing the reference level of theory for the molecule in the gas phase and produces a physically meaningful VDoS without complications.

Ideally, the model should also be able to handle pure, bulk water. To test this, we performed an NVT MD simulation of water using the MLFF model obtained from \textsc{aims-PAX}. As a further challenge, we doubled the number of atoms in the simulation box compared to the training data. No instabilities were observed during the simulation with MLFF. From the simulation we computed the RDF and compared the results obtained from an \textit{ab initio} MD simulation using the PBE functional~\cite{water_rdf_pbe}. For a detailed account of the methods used, we refer to Section~\ref{sec:comp_details}.

The RDFs are visualized in Fig.~\ref{fig:solv_results}d alongside a snapshot from the simulation. The shapes of the RDFs for both methods are very similar. For the first peak around $2.65$~\AA, a slight shift towards lower distances can be observed for the MLFF compared to the RDF obtained through DFT. Also the second peak at around 4.4~\AA is more pronounced for the RDF from the MLFF simulation. It should be noted, however, that the MLFF model was trained on PBE+MBD data, and the reference was acquired without a dispersion method. Also, PBE is known to overbind liquid water, explaining the rigidity in the water simulations.~\cite{water_rdf_pbe} Overall, it can be seen that the model is capable of handling liquid water in stable MD simulations and reproducing dynamical properties of reference calculations close to the level of theory of its training data.

Finally, the trained MLFF was tested to assess its capability to simulate an explicitly solvated paracetamol molecule. In this setup, one paracetamol molecule was immersed in a periodic box containing 600 water molecules. Notably, this system exceeds the size and complexity of all structures encountered during training, constituting a stringent extrapolation test for the MLFF.
After relaxation, multiple 800~ps long MD simulations were run. Again, more details are provided in Section~\ref{sec:comp_details}. 
No instabilities were observed throughout any of the simulations. This already hints to the fact that we are able to efficiently create a well-rounded and stable MLFF for a highly challenging system with minimal manual intervention.

To evaluate the resulting simulations, we compared the distributions of three dihedral angles, $\tau_1$, $\tau_2$, and $\tau_3$ shown in Fig.~\ref{fig:solv_results}b, in paracetamol in the gas phase and in solvent. The resulting histograms and their respective kernel density estimations are shown in Fig.~\ref{fig:solv_results}e. 
The dihedral $\tau_1$ angle describes the rotation of paracetamol's methyl group. For the gas phase, shallow minima and maxima in the distribution can be seen across all angles, meaning that the methyl group rotates freely. In contrast, in the solvent, these minima and maxima in the distribution are more pronounced, signifying a hindrance of rotation of the methyl group. This is to be expected, as the rotation does not occur in the gas phase without any obstruction but happens inside the solvation shell. The surrounding water molecules have to rearrange to accommodate the movement of the methyl group, leading to a higher energy barrier for its rotation. This ultimately leads to a reduction of its revolution frequency.

Continuing with $\tau_2$, a clear difference between the distribution in the gas and solvated phase can be observed. This angle represents the orientation of the amide in paracetamol. In the gas phase a maximum at 0$^{\circ}$ can be observed. In this case the configuration depicted in the Newman projection\cite{newman_proj} on the left in Fig.~\ref{fig:solv_results}c dominates. Here the distance between the oxygen of the amide and the hydrogen attached to carbon 3 is minimized (shown in Fig.~\ref{fig:solv_results}b as $d_{OH}$). Through this the attraction between the partial negative charge at the oxygen and the partial positive charge at the hydrogen is maximized, which has also been observed in other computational studies. \cite{Sauceda2021} In contrast, in the solvated system the distribution is broadened, and its maximum is located at 39$^{\circ}$. Its Newman projection\cite{newman_proj} is shown in Fig.~\ref{fig:solv_results}c on the right. Through this conformer, the interaction between the surrounding water molecules and the oxygen of the amide group is maximized, resulting in a lower energy state. Overall this difference between the distribution of $\tau_2$ is as expected. Whereas the intramolecular interactions are maximized in the gas phase, the equivalent holds for the intermolecular interactions in the solvated system.

Regarding $\tau_3$, for both the gas phase and paracetamol in water, a maximum of the distribution can be seen around 0$^{\circ}$. Apparently this configuration is ideal both for intramolecular interactions and interactions of solute and solvent.

To conclude the analysis of the simulation of paracetamol in explicit solvent, we want to stress that the MLFF was trained on significantly smaller systems and has not encountered these exact same geometries in the training data. Regardless, it resulted in stable MD simulations that generated physically sensible trajectories.
In summary, this application of \textsc{aims-PAX} highlights how a stable and accurate MLFF for a complex multi-species system can be created by leveraging \textsc{FHI-aims}' seamless handling of periodic and non-periodic systems.

\subsection{Computational Benefits of Parallelized  Active Learning}
\label{sec:results_bmk}

  \begin{figure}
 \centering
 \captionsetup{justification=centering}
 \includegraphics[scale=1.1]{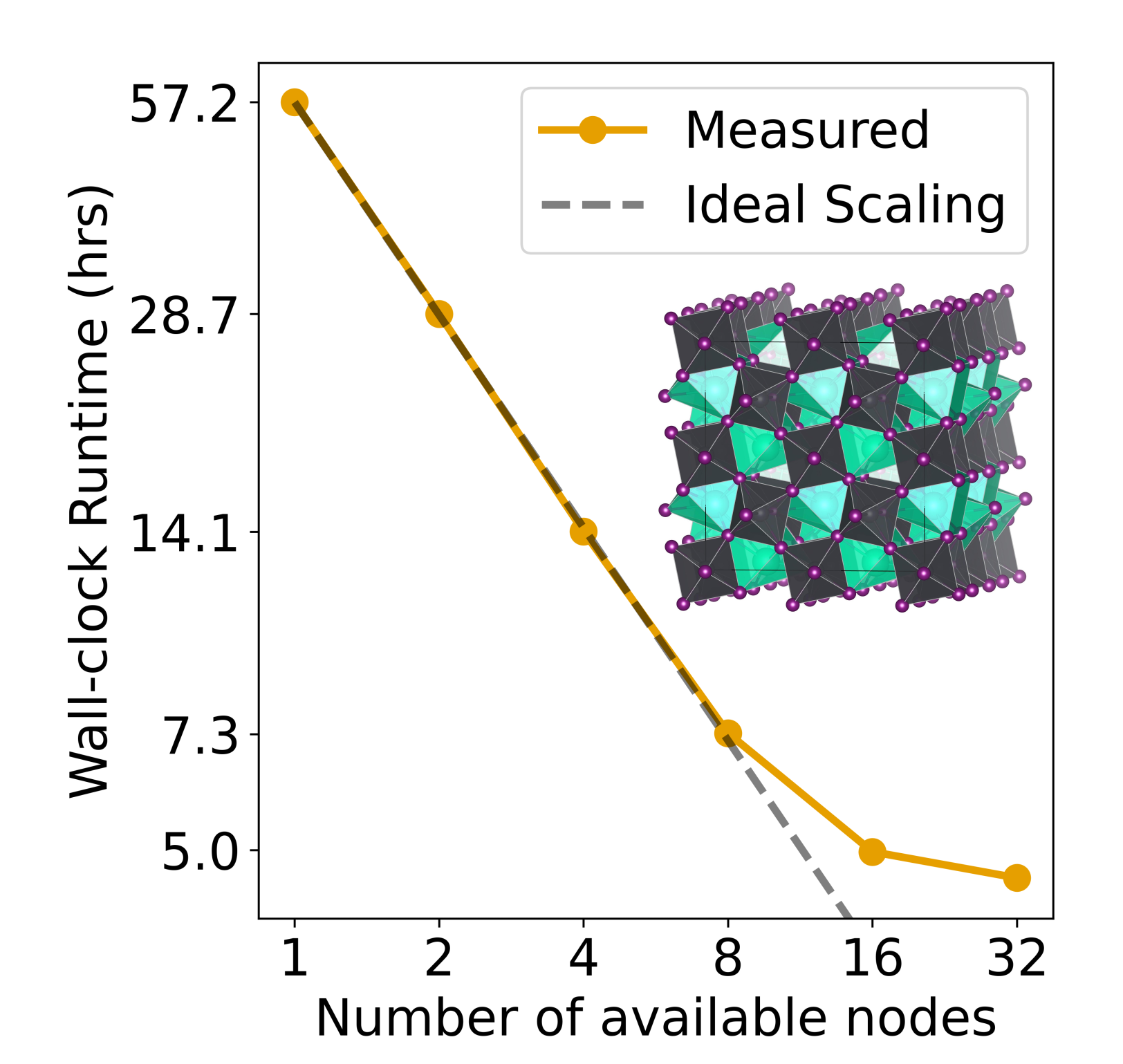}
 \caption{
 \textbf{Speedup through parallelized active learning:}
Wall-clock runtime in hours as a function of the number of available CPU nodes for \textsc{aims-PAX} using \textsc{Parsl} applied to the pervoskite CsPbI\textsubscript{3}.
 }
 \label{fig:bmk_results}
  \end{figure}

In order to investigate the efficiency of the proposed parallel AL algorithm we choose to run \textsc{aims-PAX} in parallel and serial mode for the small peptide Ac-Ac-A3-NHMe (42 atoms) and the perovskite CsPbI\textsubscript{3} (160 atoms in the unit cell). More detail regarding the exact settings for DFT and \textsc{aims-PAX} are described in Section~\ref{sec:comp_details}.

As the number of trajectories using in \textsc{aims-PAX} is integral part of the parallel algorithm, we ran the procedure with 4, 8, 16 and 32 trajectories for Ac-A3-NHMe. We used the completely serial version (the MLFF waits for DFT calculations which themselves are processed serially) and the CPU/GPU parallel version (the MLFF does not wait for DFT calculations but the latter wait for each other). The latter works through an MPI-based implementation, using ASI\cite{Stishenko2023}. The whole workflow has 1 GPU card as well as 1 CPU node with 128 cores available irrespective of how many trajectories are being used. The results of this study are shown in Section~S5.

Through the implementation using \textsc{Parsl} we can easily distribute DFT calculations across multiple nodes dynamically. Therefore, the AL becomes CPU/GPU parallel, \textit{i.e.}, DFT calculations can run while the MLFF is being used, and CPU/CPU parallel, \textit{i.e.}, multiple DFT calculations can run in parallel.
To investigate the advantage of this approach, we fix the number of trajectories to 32 and consider 1, 2, 4, 8, 16 and 32 CPU available nodes. The test is performed for the perovskite CsPbI\textsubscript{3}. We emphasize "the number of available nodes" here, as \textsc{aims-PAX} automatically scales up or down the number of workers (with one node per worker in our case) up to a user-defined maximum.

One DFT calculation, using the hardware and settings described in Section~\ref{sec:methods} for this perovskite takes around 20 user-minutes. Therefore, contrary to smaller systems where the bottleneck of the AL run can be the MLFF computation time, in the AL procedure of the perovskite the bottleneck is the DFT calculations.  This is why we chose this system to investigate the scaling of \textsc{aims-PAX} to more workers. The results of the scaling test are shown in Fig.~\ref{fig:bmk_results}. Running \textsc{aims-PAX} with only 1 available node takes about 57 hrs of wall-clock runtime. By going to two nodes, the time spent is reduced to roughly 29 h, \textit{i.e.} by a factor of nearly 2.
Doubling the number of nodes to 4 and then 8, halves the run time to 14 h and 7 h, respectively.

The deviation from ideal scaling observed when using 16 and 32 nodes, as indicated by the decreasing slope in Fig.~\ref{fig:bmk_results}, stems from the fixed number of sampling trajectories, which is 32. The likelihood that all 32 (or even 16) trajectories require halting simultaneously, and thus trigger concurrent DFT evaluations, is low. As a result, the computational resources on all nodes are not fully utilized at all times. Overall, the use of the parallel \textsc{aims-PAX} implementation can reduce MLFF creation time by a large factor for systems requiring computationally demanding reference labeling, while efficiently utilizing both GPU and CPU resources.

\section{Discussion}
In this paper, we introduced \textsc{aims-PAX}, a flexible, fully automated, open-source software package for performing AL using a parallelized algorithm that ensures efficient resource management. The key advantages of the proposed workflow include minimal human intervention, the use of state-of-the-art GP-MLFF models for initial dataset generation and pretraining, and a parallel workload manager that effectively utilizes all available computational resources. The latter is facilitated by proceeding with reference DFT calculations independently whenever a member of the MLFF ensemble is uncertain about a newly encountered geometry. The AL process is distributed across multiple sampling runs, enabling the usage of various sampling strategies, chemical species, and levels of theory.

The performance of \textsc{aims-PAX} was demonstrated on multiple challenging tasks: Ac-F-A5-K (a highly flexible peptide), a collection of diverse organic molecules, explicitly solvated paracetamol, as well as the bulk perovskite CsPbI\textsubscript{3}.

For Ac-F-A5-K, \textsc{aims-PAX} reduces the number of required DFT calculations by two orders of magnitude compared to traditional sampling approaches, providing a robust and accurate MLFF suitable for running long MD simulations. Said MLFF then models the PES of the peptide with close agreement to a model that was trained on an order of magnitude more data, underscoring the information-rich nature of the actively created data.

Furthermore, \textsc{aims-PAX} is capable of sampling data for multiple distinct chemical species at the same time. It does so by judiciously selecting structures for challenging systems more often than simpler molecules, which helps to create a balanced dataset. In the end, this led to a single, accurate MLFF that is able to run stable simulations of all molecules used during AL, highlighting the ability of \textsc{aims-PAX} to autonomously explore the chemical space of a set of molecules.

Moreover, we profited from \textsc{FHI-aims}' seamless handling of periodic and non-periodic handling and \textsc{aims-PAX}'s flexibility to create an MLFF for explicitly solvated paracetamol efficiently. We demonstrated that the resulting model could accurately and reliably simulate both bulk water and paracetamol in the gas phase, as well as in explicit water.

Finally, using the example of bulk CsPbI\textsubscript{3} perovskite, we demonstrate the advantage of parallel multi-trajectory sampling, reducing the AL time by an order of magnitude for systems requiring demanding DFT calculations.

We emphasize that the presented \textsc{aims-PAX} software package can accomplish all these tasks out of the box with minimal human effort. Strikingly, we also observed that \textsc{aims-PAX} is robust w.r.t. the choice of hyperparameters in the workflow. The default choice given in the available software enables users to swiftly apply \textsc{aims-PAX} to molecules and materials without time-consuming parametrization.

Looking ahead, the model- and \textit{ab initio}–method–agnostic design of \textsc{aims-PAX} provides a strong foundation for open-source collaboration and future innovation. We foresee its continued evolution toward greater efficiency, scalability, and accessibility. Ongoing developments aim to extend the framework through multi-GPU parallelization and optimized task scheduling, further accelerating data generation and AL cycles. Additionally, the integration of automated fine-tuning protocols for GP-MLFFs will enable the fully autonomous construction of diverse and information-rich datasets. These advances will allow \textsc{aims-PAX} to expand the accessible chemical and materials space of pre-trained MLFFs and to adapt dynamically to new systems. Beyond the selected examples presented here, current and future efforts focus on applying \textsc{aims-PAX} to increasingly complex and extended systems, bridging the gap between \textit{ab initio} accuracy and large-scale simulation realism.

\section{Methods}
\label{sec:comp_details}

DFT calculations were performed using \textsc{FHI-aims}\cite{Blum2009} version 241114 compiled as a library and called through the python ASI package \textsc{asi4py}\cite{Stishenko2023} version 1.3.18 connected with ASE\cite{HjorthLarsen2017_ASE} version 3.26.0. For \textsc{MACE}\cite{Batatia2022mace, Batatia2025design} we used \textsc{mace-torch} version 0.3.9. with \textsc{pytorch}\cite{pytorch} version 2.3.1.
For \textsc{aims-PAX} with \textsc{PARSL}, we used \textsc{PARSL} version 2024.12.16 and \textsc{FHI-aims}\cite{Blum2009} compiled as an executable. In this implementation, the ASE\cite{HjorthLarsen2017_ASE} calculator is used to perform DFT calculations.

\vskip 0.2 cm
\textbf{N-acetylphenylalanyl-pentaalanyl-lysine (Ac-F-A5-K)}
\vskip 0.2 cm

During AL, we employed the Perdew-Burke-Ernzerhof (PBE) functional\cite{PBE_Perdew1996} with non-local many-body dispersion (MBD-NL)\cite{Hermann2020_mbd_nl} using the \code{LIGHT} species defaults for numerical settings and basis sets. Relativistic corrections were applied using the atomic ZORA approximation.\cite{ZORA_vanLenthe1996} The total energy, eigenvalue, density, and force convergence criteria were set to $10^{-6}$~eV, $10^{-4}$~eV, $10^{-5}$~e/\AA$^3$, and $10^{-4}$~eV/\AA, respectively.
 For recomputing the dataset to a higher level of theory, the PBE0\cite{PBE0_Adamo1999} functional with the MBD-nl dispersion method and the \code{INTERMEDIATE} species defaults for basis sets and numerical settings was used, keeping the other settings fixed.

The serial version of \textsc{aims-PAX} was used for both IDG and AL. The former was performed by sampling 8 points for each member of an ensemble of 4 models with a stopping criterion of a maximum of 50 epochs. The structures were sampled using the small MACE-MP0 GP model\cite{batatia2024foundationmodelatomisticmaterials} by running MD in the NVT ensemble with the Langevin thermostat\cite{Berendsen1984} at 500 K with a timestep of 1~fs and a friction coefficient of $0.001~ \mathrm{fs^{-1}}$. In order to decorrelate the data points, structures were picked every 20th MD step. Their energies and forces were then computed using DFT.

The AL workflow was run until a training set size of 500 structures was reached with a 1:1 ratio for the validation set. During the AL, when new data was added to the training set, the models were trained for a total of 10 epochs on the updated dataset. The training was split into two steps, each involving 5 epochs. More precisely, this means that after 5 epochs are trained, the other running trajectories are propagated first before finishing with 5 more epochs of training. For more details on this, see Section~S5A. During AL, the structures were sampled using the same MD settings as in the IDG. The uncertainty threshold parameter $c_x$ (see Eq.~\ref{eq:threshold}) was set to the default value of 0. The uncertainty was measured every 20th MD step.

The test sets for Ac-F-A5-K were created by running MD with the small MACE-OFF\cite{Kovcs2025_maceoff} potential in the NVT ensemble at 300, 500, and 700~K using the Langevin\cite{Berendsen1984} thermostat with a friction coefficient of $0.001~ \mathrm{fs^{-1}}$ and a time step of 1~fs for 1~ns. Every 100th geometry was selected, and from the remaining points, 1000 were selected by farthest point sampling\cite{eldar1997farthest} using the mean, invariant atomic MACE-OFF descriptors. The chosen geometries were then recomputed using \textsc{FHI-aims} with the same functional and settings used in the AL.

The MD simulations for assessing the stability of models were performed in the NVT ensemble at 300, 500, and 700~K using the Langevin thermostat with a friction coefficient of $0.001~ \mathrm{fs^{-1}}$ and a time step of 1~fs for 1~ns. Throughout the simulation, the bond lengths were monitored, and if any of them exceeded 2~\AA, the simulation was stopped.\\

\vskip 0.2 cm
\textbf{MD17}
\vskip 0.2 cm

During the AL, the same settings as used in the creation of MD17\cite{chmiela2017} were used. That is, the PBE functional with the pairwise Tkatchenko-Scheffler dispersion method\cite{Tkatchenko2009} was used, employing the \code{LIGHT} species defaults for numerical settings and basis sets. The default total energy, eigenvalue, density, and force convergence criteria were used. The calculations employed a parallel KS method with load balancing. 

The parallel version of \textsc{aims-PAX} employing \textsc{Parsl} was used for both IDG and AL. The former was performed by sampling 10 points for each member of an ensemble of 4 models and for each distinct chemical species. After sampling, 5 training epochs were performed before continuing sampling up to a stopping criterion of 20 epochs. The structures were sampled using the small MACE-MP0 GP model\cite{batatia2024foundationmodelatomisticmaterials} at 500 K with a timestep of 1~fs and a friction coefficient of $0.001~ \mathrm{fs^{-1}}$. In order to decorrelate the data points, structures were picked every 25th MD step. Their energies and forces were then computed using DFT.

The AL workflow was run until a training set size of 1000 structures was reached with a 10:1 ratio for the validation set. During the AL, when new data was added to the training set, the models were trained for a total of 1 epoch on the updated dataset. During AL, the structures were sampled using the same MD settings as in the IDG. The uncertainty threshold parameter $c_x$ (see Eq.~\ref{eq:threshold}) was set to the default value of 0. The threshold was frozen after the training set size reached a size of 500 geometries. The uncertainty was measured every 25th MD step.

The data for training the MLFF from scratch on MD17 was obtained by taking the first 50k structures for each molecule in the original dataset. Then every 25th structure was selected from this subset, resulting in 2,000 points per species. From these 2,000 points, 100 points for training and 10 points for validation were randomly chosen per species. All of these subsets were then combined, resulting in 1,000 training and 100 validation points.\\

\begin{table*}
\centering
\begin{tabular}{|c|c c c c|}
\cline{2-5}
\multicolumn{1}{c|}{} & \multicolumn{4}{c|}{\textbf{System}} \\
\hline
\textbf{Parameter} & \textbf{Ac-F-A5-K (small) / MD17} &
\textbf{Ac-F-A5-K (large)} & \textbf{CsPbI\textsubscript{3}} & \textbf{Paracetamol+H\textsubscript{2}O} \\
\hline
Channels & 32 & 256 & 64 & 128\\
\hline
Max degree $L_{\text{max}}$ & 1  & 2 & 1 & 1 \\
\hline
Cutoff [\r{A}] & 6  & 6 & 6 & 6 \\
\hline
Radial Bessel functions & 8 & 8 & 10 & 8 \\
\hline
Message-passing layers & 2 & 2 & 2 & 2 \\
\hline
Correlation order & 3 & 3 & 3 & 3 \\
\hline
Radial MLP layers & 3 & 3 & 3 & 3 \\
\hline
Neurons per MLP layer & 32 & 64 & 16 & 64 \\
\hline
Activation function & SiLU & SiLU & SiLU & SiLU \\
\hline
Output Layer Irreps & "128x0e" & "16x0e" & "16x0e" & "128x0e" \\
\hline
\end{tabular}
\caption{
Architectural parameters of the MACE models used for the systems studied in this work.
}
\label{tab:mace_architectures}
\end{table*}
\vskip 0.2 cm
\textbf{Paracetamol in Water}
\vskip 0.2 cm

During the AL, the PBE functional with many-body dispersion (MBD)\cite{Tkatchenko2009} using the \code{LIGHT} species defaults for numerical settings and basis sets was used. The default total energy, eigenvalue, density, and force convergence criteria were used. The calculations employed a parallel KS method with load balancing. For the periodic system a k grid of $2\times2\times2$ was utilized.

The initial structures for paracetamol surrounded by a cluster of 90 water molecules, bulk water with 64 and 128 molecules at a density of 1 g/mL and paracetamol surrounded by 600 water molecules at a density of 1 g/mL were created using \textsc{Packmol}\cite{packmol} version 21.1.0.

Paracetamol in the gas phase, bulk water with 64 molecules, and paracetamol surrounded by a water cluster were optimized using \textsc{FHI-aims} with a convergence threshold on the force of 0.01 eV/\AA ~ before using them in \textsc{aims-PAX}. Before MD production runs of paracetamol in the gas phase, bulk water with 128 molecules, and paracetamol in explicit water (600 water molecules), the  structures were first optimized using \textsc{FHI-aims} for the former and the medium MACE-MP0 GP model\cite{batatia2024foundationmodelatomisticmaterials} for the latter two with  a convergence threshold on the force of 0.01 eV/\AA. Then optimization was repeated using the MACE model acquired from the \textsc{aims-PAX} run with a convergence threshold on the force of 0.01 eV/\AA. All optimizations were performed using the Broyden–Fletcher–Goldfarb–Shanno algorithm\cite{Fletcher2000} as implemented in \textsc{FHI-aims} and ASE\cite{HjorthLarsen2017_ASE}, respectively.

The parallel version of \textsc{aims-PAX} employing \textsc{Parsl} was used for both IDG and AL. The former was performed by sampling 5 points for each member of an ensemble of 4 models and for each trajectory. After sampling, 5
training epochs were performed before continuing sampling up to a stopping criterion of 20 epochs. The structures were
sampled using the medium MACE-MP0 GP model\cite{batatia2024foundationmodelatomisticmaterials} from 7 trajectories in total. Of these, one was of paracetamol in the gas phase at 300 K (NVT) with a timestep of 1 fs and a friction coefficient of 0.001 fs$^{-1}$; three were of paracetamol surrounded by a cluster of 90 water molecules at 300~K, 350~K, and 400~K (NVT) with a timestep of 0.5 fs and a friction coefficient of 0.001 fs$^{-1}$; and the remaining three were of bulk water at 1~atm (NPT) and 300~K, 400~K, and 500~K with a timestep of 0.5 fs using full Martyna-Tobias-Klein (MTK) dynamics\cite{Martyna1994} as implemented in ASE\cite{HjorthLarsen2017_ASE} with a temperature and pressure damping factor of 100 and 1000, respectively. In order to decorrelate the data points, structures were picked every 25th MD step. Their energies and forces were then computed using DFT.

The AL workflow was run until a training set size of 1,000 structures was reached with a 10:1 ratio for the validation set. During the AL, when new data was added to the training set, the models were trained for a total of 1 epoch on the updated dataset. During AL, the structures were sampled using the same MD settings as in the IDG. The uncertainty threshold parameter $c_x$ (see Eq.~\ref{eq:threshold}) was set to the default value of 0 and the uncertainty was measured every 50th MD step. The threshold was frozen after the training set size reached a size of 500 geometries.

In order to acquire the vibrational density of states (VDoS) of paracetamol in the gas phase, the following protocol was applied. First, 20 starting geometries were picked through k-means clustering from the MD17 dataset using the dihedral angles $\tau_1, \tau_2, \tau_3$ (see Fig.~4b) as the descriptor. From the 20 clusters, the geometry closest to the respective cluster centers was used as a starting point for a 10 ps NVT simulation at 300~K with a timestep of 1~fs and a friction coefficient of 0.001 fs$^{-1}$. The final structures and their velocities were then used to perform MD runs in the NVE ensemble with a timestep of 0.1 fs. Subsequently, the velocity autocorrelation function (ACF) was computed from the combined trajectories, and the VDoS was acquired through a Fourier transform of the velocity ACF.

The oxygen-oxygen radial distribution function (RDF) of bulk water was acquired from a 500 ps MD simulation of 128 water molecules with periodic boundary conditions, whereas the first 200 ps were used to equilibrate the system. The simulation was performed in the NPT ensemble using MTK dynamics at 330K and 1 atm. The RDF was computed using the analysis tools implemented in ASE\cite{HjorthLarsen2017_ASE}.

In order to acquire the distributions of dihedral angles $\tau_1, \tau_2, \tau_3$ (see Fig.~4b) of paracetamol in the gas phase, 20 independent 800 ps long MD simulations in the NVT ensemble from the optimized geometry at 300~K with a timestep of 0.5~fs and a friction coefficient of 0.001 fs$^{-1}$ were performed. The first 50 ps were used for equilibration.
In case of paracetamol in water, 36 independent 800 ps long MD simulations in the NVT ensemble at 300~K with a timestep of 0.5~fs and a friction coefficient of 0.001 fs$^{-1}$ were performed. The starting geometries were chosen from a separate MD run so that the dihedral angles $\tau_2$ and $\tau_3$ were equally distributed. The first 50 ps were used for equilibration.

\vskip 0.2 cm
\textbf{Bulk Perovskite (CsPbI\textsubscript{3} 2x2x2)}
\vskip 0.2 cm

During the AL, the PBE functional with the pairwise Tkatchenko-Scheffler dispersion method\cite{Tkatchenko2009} was used, employing the \code{INTERMEDIATE} species defaults for numerical settings and basis sets. Relativistic corrections were applied using the atomic ZORA approximation.\cite{ZORA_vanLenthe1996}
The total energy, eigenvalue, density, and force convergence criteria were set to $10^{-6}$~eV, $10^{-5}$~eV, $10^{-5}$~e/\AA$^3$, and $10^{-4}$~eV/\AA, respectively.
A Gaussian smearing of $0.05$~eV was applied to the orbital occupations. The calculations employed a parallel KS method with load balancing and local indexing enabled. A maximum of $300$ self-consistency iterations was allowed. The charge mixing parameter was set to $0.02$. The k grid was set to $1\times1\times1$. The lattice vectors were [17.23958, 0, 0], [0, 17.23958, 0], and [0, 0, 25.00256], all in \AA ngstrom.

The parallel version of \textsc{aims-PAX} employing \textsc{Parsl} was used for both IDG and AL. The former was performed by sampling 10 points for each member of an ensemble of 4 models with a stopping criterion of 50 epochs for the initial training. The structures were sampled using the small MACE-MP0 GP model\cite{batatia2024foundationmodelatomisticmaterials} by running NPT MD with the Nosé-Hoover thermostat\cite{Evans1985_NH_thermo} at 300~K and the Parinello-Rahman barostat\cite{Parrinello1981_NPT} at 1~bar. The timestep was set to 1 fs and otherwise default ASE\cite{HjorthLarsen2017_ASE} parameters were used. In order to decorrelate the sampled data points, structures were picked every 20th MD step. Their energies, forces, and stress were then computed using DFT. The models were converged on the initial dataset before continuing with AL by training them until no improvements w.r.t. the validation set were achieved for 50 epochs 

The AL workflow was run until a training set size of 100 structures was reached with a 7:3 ratio for the validation set. The maximum epochs per trajectory were 10, and the intermediate epochs were 10. The structures were sampled using the same MD settings as described in the IDG above. The uncertainty threshold parameter $c_x$ (see Eq.~\ref{eq:threshold}) was set to 0.2, and the uncertainty itself was checked every 10th MD step.

\vskip 0.2 cm
\textbf{Settings for MACE}
\vskip 0.2 cm
The MACE architectures used during the \textsc{aims-PAX} runs are summarized in Table~\ref{tab:mace_architectures}.

For the training during AL with \textsc{aims-PAX} the following settings were used. The AMSGrad optimizer \cite{reddi2018amsgrad} and a learning rate of 0.01 were utilized throughout. For the IDG, the learning rate was decreased by 0.8 using the \textit{Reduce On Plateau} scheduler with a patience of 5 and $\gamma = 0.9993$. No learning rate scheduler was used during the AL. An exponential moving average of 0.99 for the model parameters and a gradient clipping of 10 were used.
For the loss function, a weighted mean square loss of energies and forces with weights 1 and 1000, respectively, was utilized. A batch size of 5 was used throughout.
After the AL runs themselves, the best-performing models of the respective ensembles were trained on the final training set until there was no improvement w.r.t. the validation set for 50 epochs.

For training the large model for Ac-F-A5-K (third column in Table~\ref{tab:mace_architectures}) from scratch on the recomputed dataset, the same settings as described above for \textsc{aims-PAX} were used except for the following changes. The energy and force weights of the loss function were set to 44 and 1000, respectively. The model was trained for 1000 epochs, and after 750 epochs, the energy and force weights were swapped and the learning rate set to 0.001.
The learning rate was decreased by 0.8 using the \textit{Reduce On Plateau} scheduler with a patience of 256 and $\gamma = 0.9993$.

For training the model for MD17 from scratch the same settings as described above for \textsc{aims-PAX} were used except for the following changes. The model was trained for a total of 500 epochs. The MACE architecture was the one summarized in the first column in Table~\ref{tab:mace_architectures}.

\vskip 0.2 cm
\textbf{Hardware}
\vskip 0.2 cm

The benchmarks, the active learning, and the training were performed using an NVIDIA Ampere 40 GB HBM GPU and an AMD EPYC Rome 7452 CPU. Recomputing the data on a higher level of theory and the DFT calculations through \textsc{Parsl} were done using AMD EPYC Rome 7H12. The training of MACE models from scratch and MD runs for Ac-F-A5-K were performed using an NVIDIA A100 80GB GPU.

\bibliography{references}

\begin{acknowledgments}
T.H. thanks Sander Vandenhaute for his time and valuable suggestions concerning the \textsc{Psiflow} and \textsc{Parsl} packages. The authors also thank Dr. Iryna Knysh for her support in debugging the code and Sergio Suárez for support in calculating vibrational spectra.\\
\indent T.H. acknowledges financial support from the Luxembourg National Research (FNR) under the AFR project 17932705. T.H. and A.T. acknowledge financial support from Molecular Simulations from First Principles e.V. (MS1P). I.P. and A.T. acknowledge the Luxembourg National Research Fund under grant FNR-CORE MBD-in-BMD (18093472) and the European Research Council under ERC-AdG grant FITMOL (101054629).
The simulations were performed on the HPC facilities of the University of Luxembourg 
(see \href{hpc.uni.lu}{hpc.uni.lu}), the Luxembourg national supercomputer MeluXina, the computing resources at the Max Planck Institute for the Stucture and Dynamics of Matter in Hamburg and at the MPCDF. The authors gratefully acknowledge the LuxProvide teams for their expert support.  S.S. acknowledges support from the UFAST International Max Planck Research School.
\end{acknowledgments}
\section*{Author Contributions}
 T.H. conceptualization, investigation, data curation, formal analysis, funding acquisition, methodology, software, validation, visualization, writing – original draft, writing – review \& editing

S.S. support in validation and data analysis, writing - review \& editing

M.R. conceptualization, formal analysis, funding acquisition, methodology, supervision, project administration, writing – review \& editing

A.T. conceptualization, formal analysis, funding acquisition, methodology, writing – review \& editing

I.P. conceptualization, formal analysis, funding acquisition, methodology, supervision, project administration, writing – original draft, writing – review \& editing

\section*{Data Availability Statement}
The data and models used for the results in this study can be found on: \href{https://doi.org/10.5281/zenodo.17359257}{10.5281/zenodo.17359257}

The code of \textsc{aims PAX} is available under the \textsc{Github} repository: \href{https://github.com/tohenkes/aims-PAX}{github.com/tohenkes/aims-PAX}.



\end{document}


\preprint{AIP/123-QED}

\title[SI: aims-PAX: Parallel Active eXploration Enables Expedited Construction of Machine Learning Force Fields for Molecules and Materials]{SI: aims-PAX: Parallel Active eXploration Enables Expedited Construction of Machine Learning Force Fields for Molecules and Materials}
\author{Tobias Henkes}
\affiliation{ 
Department of Physics and Materials Science, University of Luxembourg, L-1511 Luxembourg, Luxembourg
}
\author{Shubham Sharma}
\affiliation{%
Max Planck Institute for the Structure and Dynamics of Matter, 22761 Hamburg, Germany
}%
\author{Alexandre Tkatchenko}%
\affiliation{ 
Department of Physics and Materials Science, University of Luxembourg, L-1511 Luxembourg, Luxembourg
}%
\author{Mariana Rossi}
\affiliation{%
Max Planck Institute for the Structure and Dynamics of Matter, 22761 Hamburg, Germany
}%
\author{Igor Poltavskyi}%
\email{igor.poltavskyi@uni.lu}
\affiliation{ 
Department of Physics and Materials Science, University of Luxembourg, L-1511 Luxembourg, Luxembourg
}

\maketitle

\clearpage
\section*{Supplementary Information}

\beginsupplement
\section{Accuracy and Stability of the models for Ac-F-A5-K}

The average MAEs for different numbers of concurrent trajectories during the AL procedure are reported along with their variances in Table~\ref{tab:peptide_accuracy}. At 300~K, all models achieved MAEs between 16 and 18~$\mathrm{meV}/\mathrm{\text{\r{A}}}$, independent of the number of trajectories used. Similarly, for the 500 and 700~K test sets, MAEs ranged from 27 to 30 and 37 to 40~$\mathrm{meV}/\mathrm{\text{\r{A}}}$, respectively; showing negligible dependence on the number of parallel trajectories. \\

\begin{table}[h]
\setlength{\tabcolsep}{12pt} 
\begin{tabular}{c|ccc|}
\cline{2-4}
\multicolumn{1}{l|}{}                          & \multicolumn{3}{c|}{\textbf{MAE Forces (meV/\AA)}}                                                             \\ \hline
\multicolumn{1}{|c|}{\textbf{\# Traj.}} & 300 K                          & 500 K                          & 700 K                          \\ \hline
\multicolumn{1}{|c|}{1}                        & 17.93~$\pm$~2.28               & 27.07~$\pm$~0.25               & 37.73~$\pm$~0.84               \\ \hline
\multicolumn{1}{|c|}{4}                        & 16.60~$\pm$~0.44               & 28.50~$\pm$~0.61               & 37.47~$\pm$~1.11               \\ \hline
\multicolumn{1}{|c|}{8}                        & 16.43~$\pm$~0.67               & 28.73~$\pm$~0.99               & 37.93~$\pm$~1.36               \\ \hline
\multicolumn{1}{|c|}{16}                       & 17.20~$\pm$~0.82               & 29.67~$\pm$~0.95               & 39.70~$\pm$~1.77               \\ \hline
\multicolumn{1}{|c|}{32}                       & 16.83~$\pm$~0.38               & 29.77~$\pm$~1.30               & 38.73~$\pm$~1.36               \\ \hline
\end{tabular}
\caption{\textbf{Mean absolute test errors of models of Ac-F-A5-K acquired using \textsc{aims-PAX}:} Models were created from various number of sampling trajectories (\# Traj.). Average and standard deviation over the best models from three separate \textsc{aims-PAX} runs.}
\label{tab:peptide_accuracy}
\end{table}

The results of all MD stability tests are summarized in Table~\ref{tab:peptide_stability}. At 300~K, all MD trajectories remained stable, regardless of the number of trajectories used during the AL process. At 500~K, the temperature used during AL, nearly all MD runs were also stable, with only one exception: a single unstable simulation was observed for the model trained using eight parallel AL trajectories. In contrast, at the elevated temperature of 700~K, instability was observed in at least one MD run for every MLFF model tested. Specifically, for models trained with 8 and 32 trajectories, 7 out of 12 MD simulations were unstable. For the model obtained from a single-trajectory AL run, 3 simulations were unstable. Finally, for models trained with 4 and 16 trajectories, 1 and 2 simulations were unstable, respectively. These results align with the sampling strategy used in AL: since the training data were collected at 500~K, it is expected that MD simulations at or below this temperature (e.g., 300~K and 500~K) remain stable, as the MLFF is unlikely to encounter configurations outside its training domain. At 700~K, however, the MD trajectories explore more diverse and potentially unseen regions of configuration space, which can lead to instability due to extrapolation beyond the model’s training domain.

\begin{table}[h]
\setlength{\tabcolsep}{12pt}
\begin{tabular}{c|ccc|}
\cline{2-4}
\multicolumn{1}{l|}{}                          & \multicolumn{3}{c|}{\textbf{\# Stable MD Runs}}                                     \\ \hline
\multicolumn{1}{|c|}{\textbf{\# Traj.}}        & 300 K & 500 K & 700 K \\ \hline
\multicolumn{1}{|c|}{1}                        & 12    & 12    & 9     \\ \hline
\multicolumn{1}{|c|}{4}                        & 12    & 12    & 11    \\ \hline
\multicolumn{1}{|c|}{8}                        & 12    & 11    & 7     \\ \hline
\multicolumn{1}{|c|}{16}                       & 12    & 12    & 10    \\ \hline
\multicolumn{1}{|c|}{32}                       & 12    & 12    & 7     \\ \hline
\end{tabular}
\caption{
\textbf{Number of stable MD simulations performed with models of Ac-F-A5-K acquired using \textsc{aims-PAX}:} Various number of trajectories (\# Traj.) were used for sampling at multiple temperatures. Three models were obtained from separate \textsc{aims-PAX} runs and four simulations were run for each, thus a maximum of 12 stable MD runs can be achieved per category. Stability was defined as no bonded atoms separated by more than 2 \AA.
}
\label{tab:peptide_stability}
\end{table}

\section{Training models from scratch}
\label{sec:train_fr_scratch}
\begin{table}[h]
\centering
\begin{tabular}{|c|c|c|}
\hline
\textbf{Temperature} & \textbf{MAE Forces (meV/\AA)} & \textbf{\# Stable MD Runs} \\ \hline
300 K                & 17.53 $\pm$ 0.90                             & 12                          \\ \hline
500 K                & 28.03 $\pm$ 1.03                             & 12                          \\ \hline
700 K                & 37.77 $\pm$1.25                             & 10                          \\ \hline
\end{tabular}
\caption{
\textbf{Stability and test errors for models of Ac-F-A5-K trained from scratch:}
Mean absolute test errors (mean $\pm$ standard deviation) and number of stable MD simulations at various temperatures for models trained from scratch on data acquired via an \textsc{aims-PAX} run. MAE values are averaged over three models with different seeds. Each model was used to generate 4 MD simulations, for a total of 12 per temperature. Stability was defined as no bonded atoms separated by more than 2 \AA.
}
\label{tab:train_fr_scratch}
\end{table}

In order to assess the difference between continuously training the models during AL and training models from scratch afterwards, we trained 3 models with different seeds on the dataset of Ac-F-A5-K, acquired through the \textsc{aims-PAX} run with 4 trajectories as described in Section~IV. All settings for training, testing, MD simulations and the model architecture were kept the same. The results for the accuracy and stability are shown in Table~\ref{tab:train_fr_scratch}. Comparing with results in Tables~SI and SII, there is no meaningful difference between training from scratch and using continual learning (CL) observable for accuracy and stability. Given that CL is computational more efficient, it is the mode of action used in \textsc{aims-PAX}.\\

\section{Accuracy of the models for MD17}
\label{sec:md17_test_error}

  \begin{figure*}
 \centering
 \captionsetup{justification=centering}
 \includegraphics[scale=0.5]{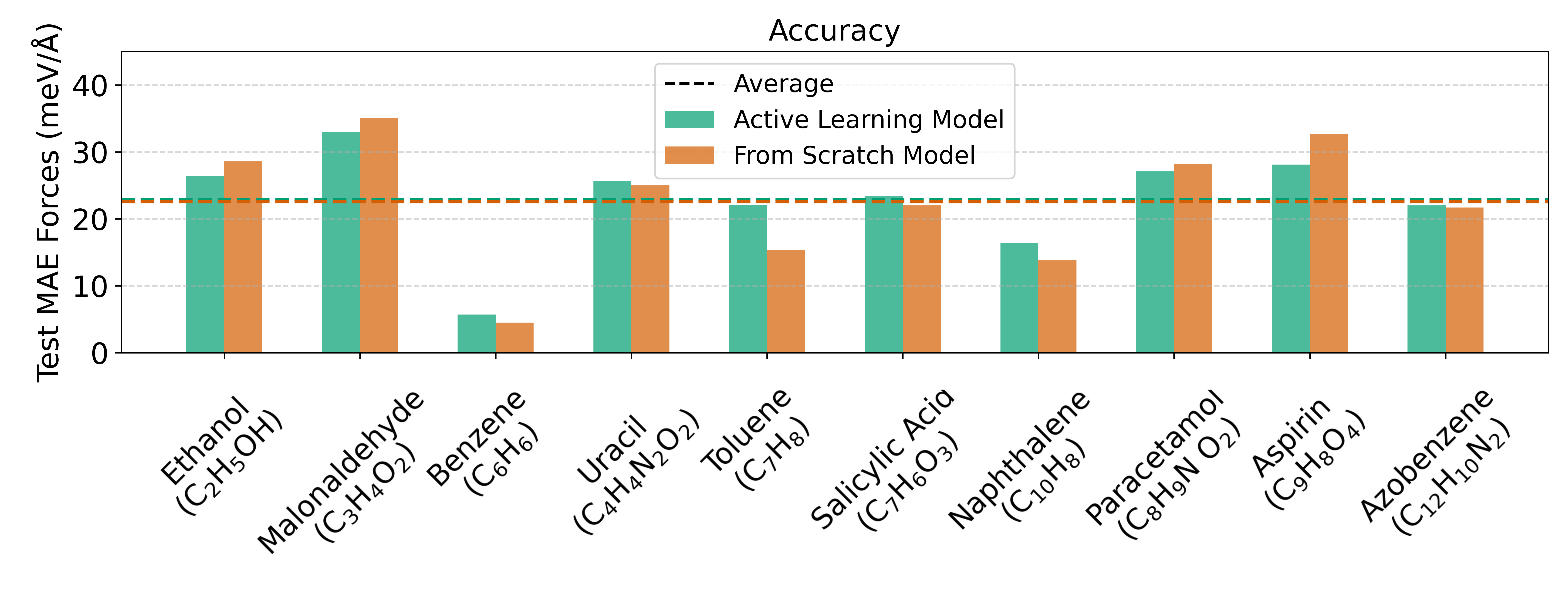}
 \caption{
\textbf{Accuracy of MLFFs acquired using \textsc{aims-PAX} and a model trained from a manually created dataset for MD17:} The dashed line indicates average performance across all systems. Systems are sorted from the smallest to the largest number of atoms in the molecule}
 \label{fig:md17_results_si}
  \end{figure*}

The test errors of the MLFFs trained using \textsc{aims-PAX} and a manual, "traditional" approach are shown in Fig.~\ref{fig:md17_results_si}.
The dashed line shows the MAE on the forces across all species, and the bars show system-specific MAEs. Both the model trained from scratch and the model generated using \textsc{aims-PAX} perform similarly with an overall MAE of 22.6 and 22.9 meV/\AA, respectively. In particular, for benzene and naphthalene, both models achieved low errors of 5.7 and 4.5 meV/\AA, as well as 16.4 and 13.8 meV/\AA, for \textsc{aims-PAX} and the model traditionally trained, respectively.

The model from the AL procedure is only slightly less accurate, while trained on only 33 and 39 points for benzene and naphthalene, respectively, compared to 100 points each for the reference model. Similarly, the model from \textsc{aims-PAX} was trained on only 39 toluene geometries and achieved an error of 22.1 meV/\AA, while the traditional model was trained on 100 geometries and achieved an error of 13.8 meV/\AA.

Both model types exhibit their largest errors on aspirin and malonaldehyde, with MAEs of 32.7 and 35.1 meV/\AA, respectively, for the model created manually and 28.1 meV/\AA~for aspirin and 33.0 meV/\AA~for malonaldehyde for the model obtained \textit{via} \textsc{aims-PAX}.

\section{Speedup of CPU/GPU parallel aims-PAX version}
\label{sec:speedup_peptide}

\begin{figure}
 \centering
 \captionsetup{justification=centering}
 \includegraphics[scale=1.1]{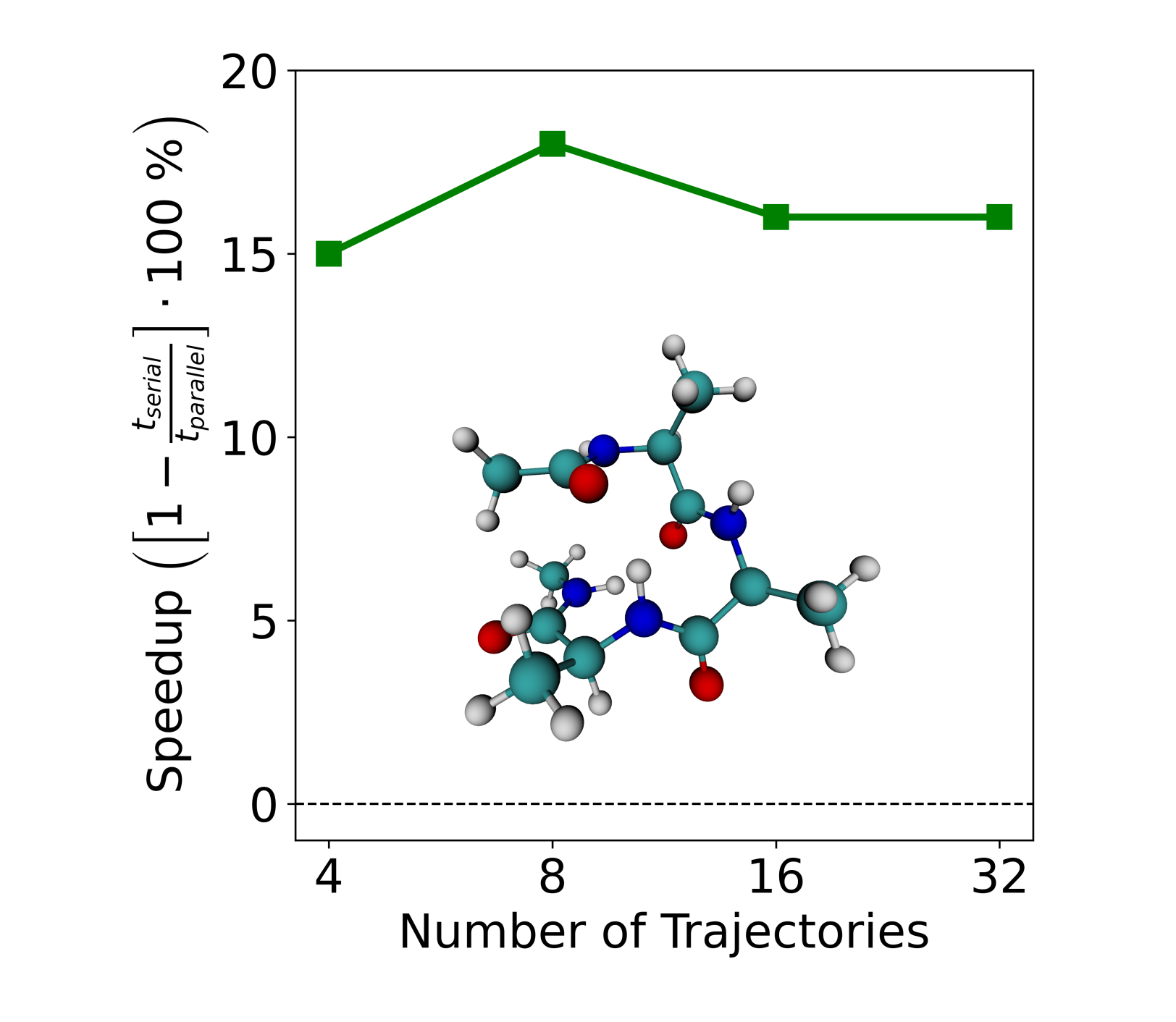}
 \caption{
\textbf{ Speedup of the parallel version over the serial version:} Application to Ac-Ala3-NHMe running on a single CPU node with 128 cores and 1 GPU card.
 }
 \label{fig:bmk_results_peptide}
  \end{figure}
The results regarding the speedup using the CPU/GPU parallel \textsc{aims-PAX} version are summarized in Fig.~\ref{fig:bmk_results_peptide}. We found that using 4 or more trajectories results in a speedup of 15~\%. The largest speed-up can be observed with 8 trajectories with 18~\%. For 16 and 32 the speedup is slightly lower than for 8 trajectories at 16~\% for both runs. 
The speed-up observed when enabling more trajectories is caused by DFT calculations that run concurrently with the sampling of new points and training. Furthermore, a slight speed-up is observed between 4 and 8 trajectories, likely due to the reduced probability of no trajectory being propagated at a given time. Because trajectories are stopped when the uncertainty threshold is exceeded, it is more likely that all trajectories will stop if there are fewer of them, thus halting all AL progress.

The same DFT settings used for Ac-F-A5-K were used for Ac-A3-NHMe during AL. The only exception is that no dispersion correction was applied.
For generating the initial data set using \textsc{aims-PAX} the same settings as for Ac-F-A5-K were used.

For the AL workflow with \textsc{aims-PAX}, the same settings were used as for Ac-F-A5-K, except that the parallel \textsc{aims-PAX} version was used and the procedure stopped when the training set size of 200 was reached. Also, the same MACE architecture as described in Table~I under \textit{Ac-F-A5-K (small)} was used.

\section{Technical Details of \textsc{aims-PAX}}
Here we provide some technical details of the inner workings of \textsc{aims-PAX}. We focus on the training during AL, explain how data as well as failed SCF convergence are handled and highlight slight differences between the order of operations in the serial and parallel AL algorithm.
\label{sec:tech_details_algo}

\subsection{Training during Initial Dataset Generation and Active Learning}
\label{sec:train_idg_al}
During the initial dataset generation, the user decides how many points are selected for each ensemble member during each sampling step. For example, the user specifies that 5 structures per member are to be selected. \textsc{aims-PAX} then runs the sampling algorithm, and once 5 points are picked from the trajectory for each member (and labels are computed), the models are trained for a user specified number of epochs, namely \code{intermediate_epochs}. This process is repeated multiple times \textit{i.e.} running sampling, picking points, labeling and training. The rationale behind this is, that if the user wants to have a specific accuracy of the models before running AL, \textsc{aims-PAX} makes sure that not too many structures are sampled and DFT calculations are performed.

During the AL procedure each trajectory is associated with a state. Technically speaking, a loop is performed over all trajectories and depending on their state, different actions are performed. At the beginning of AL, all trajectories have the state \code{running}, which means the sampling algorithm is performed. Once a point is picked for labeling, the state of this specific trajectory is set to \code{waiting} until the DFT calculation is done and the results were received. This then changes the state to \code{training} and the user specified number of training epochs are performed, these are called \code{intermediate_epochs_al}. Afterwards, \textsc{aims-PAX} continues the loop over the trajectories. Only once a maximum, user-specified number of epochs, \code{  max_epochs_worker}, is reached, the trajectory's state swtiches back to \code{running}. This is done to enable other trajectories to continue sampling, potentially triggering new DFT calculations, which can then run while the models are trained. In addition, this means that the trajectories are always propagated with continuously updated model parameters.
\subsection{Resetting the Optimizer during Active Learning}
While training during AL, the weights are not reset when a new point is added. We have seen that repeatedly using the updated weights can result in the model being stuck in a minimum. We found it advantageous in this case to reset the optimizer state if the model is not improved after \code{max_epochs_worker} epochs (see Section~\ref{sec:train_idg_al}). This deletes the history of the adaptive optimizer (e.g. Adam or AMSGrad), resulting in a larger learning rate which helps the model to leave the local minimum.
\subsection{Handling of new data points during Active Learning}
Once a new point is selected and labeled during AL, it is either added in the training or validation set. To which dataset the new point contributes depends on a user specified ratio, that is kept consistent, e.g. 0.5, which means points are added to both sets alternately. In contrast to the IDG, both datasets are shared across models (except for the initial starting points that are present before the AL). 
\subsection{Handling of Failed SCF Convergence}
During the IDG, if a DFT computation does not converge the geometry is discarded from the dataset. The procedure then just continues until any stopping criterion is met. However, we have not noticed any instances where SCF convergence could not be achieved for a geometry generated by the GP model for our systems.\\
In the case of AL, points where the SCF cycles do not converge are also discarded. On the trajectory where this is the case, a checkpoint geometry is loaded. This checkpoint is updated each time a selected structure is successfully labeled using DFT and the data is added to the training set. This ensures, that if the checkpoint is loaded, the MD continues from a geometry that is known to the MLFF.
\subsection{Operational Differences: Serial vs. Parallel Version}
While the overall AL workflow of \textsc{aims-PAX} is the same for its serial and parallel versions, there are slight differences that we want to point out. In the case of the serial procedure, the sampling and training of the ML models is halted if DFT calculations are performed. Afterwards, the model parameters are updated on the new data. Practically this results in all trajectories being propagated with the new information. For the parallel version, other trajectories can be propagated during the DFT calculation, meaning that sampling is done without the information of the current DFT calculation. While it can mean that potentially redundant points are sampled, the computational benefit,  and thus possibility of scaling up the workflow, outweighs this inefficiency.
